\documentstyle[referee]{mn}
\newif\ifAMStwofonts

\def\beq{\begin{equation}}\def\eeq{\end{equation}}
\def\n{n}
\def\p{p}

\def\np{{\p}}
\def\d{\delta}
\def\K{{\cal K}}
\def\mun{{\mu_n}}
\def\mup{{\mu_p}}

\ifoldfss
  \ifCUPmtlplainloaded \else
    \NewTextAlphabet{textbfit} {cmbxti10} {}
    \NewTextAlphabet{textbfss} {cmssbx10} {}
    \NewMathAlphabet{mathbfit} {cmbxti10} {} 
    \NewMathAlphabet{mathbfss} {cmssbx10} {} 
  \fi
  \ifAMStwofonts
    \ifCUPmtlplainloaded \else
      \NewSymbolFont{upmath} {eurm10}
      \NewSymbolFont{AMSa} {msam10}
      \NewMathSymbol{\upi}     {0}{upmath}{19}
      \NewMathSymbol{\umu}     {0}{upmath}{16}
      \NewMathSymbol{\upartial}{0}{upmath}{40}
      \NewMathSymbol{\leqslant}{3}{AMSa}{36}
      \NewMathSymbol{\geqslant}{3}{AMSa}{3E}

      \let\leq=\leqslant 
       
    \fi
  \fi
\fi 

\ifnfssone
  \newmathalphabet{\mathit}
  \addtoversion{normal}{\mathit}{cmr}{m}{it}
  \addtoversion{bold}{\mathit}{cmr}{bx}{it}
  \newmathalphabet{\mathbfit} 
  \addtoversion{normal}{\mathbfit}{cmr}{bx}{it}
  \addtoversion{bold}{\mathbfit}{cmr}{bx}{it}
  \newmathalphabet{\mathbfss} 
  \addtoversion{normal}{\mathbfss}{cmss}{bx}{n}
  \addtoversion{bold}{\mathbfss}{cmss}{bx}{n}
  \ifAMStwofonts
    \ifCUPmtlplainloaded \else
      \UseAMStwoboldmath
      \makeatletter
      \new@mathgroup\upmath@group
      \define@mathgroup\mv@normal\upmath@group{eur}{m}{n}
      \define@mathgroup\mv@bold\upmath@group{eur}{b}{n}
      \edef\UPM{\hexnumber\upmath@group}
      \new@mathgroup\amsa@group
      \define@mathgroup\mv@normal\amsa@group{msa}{m}{n}
      \define@mathgroup\mv@bold\amsa@group{msa}{m}{n}
      \edef\AMSa{\hexnumber\amsa@group}
      \makeatother
      \mathchardef\upi="0\UPM19
      \mathchardef\umu="0\UPM16
      \mathchardef\upartial="0\UPM40
      \mathchardef\leqslant="3\AMSa36
      \mathchardef\geqslant="3\AMSa3E

      \let\leq=\leqslant 

    \fi
  \fi
\fi 

\ifnfsstwo
  \DeclareMathAlphabet{\mathbfit}{OT1}{cmr}{bx}{it}
  \SetMathAlphabet\mathbfit{bold}{OT1}{cmr}{bx}{it}
  \DeclareMathAlphabet{\mathbfss}{OT1}{cmss}{bx}{n}
  \SetMathAlphabet\mathbfss{bold}{OT1}{cmss}{bx}{n}
  \ifAMStwofonts
    \ifCUPmtlplainloaded \else
      \DeclareSymbolFont{UPM}{U}{eur}{m}{n}
      \SetSymbolFont{UPM}{bold}{U}{eur}{b}{n}
      \DeclareSymbolFont{AMSa}{U}{msa}{m}{n}
      \DeclareMathSymbol{\upi}{0}{UPM}{"19}
      \DeclareMathSymbol{\umu}{0}{UPM}{"16}
      \DeclareMathSymbol{\upartial}{0}{UPM}{"40}
      \DeclareMathSymbol{\leqslant}{3}{AMSa}{"36}
      \DeclareMathSymbol{\geqslant}{3}{AMSa}{"3E}

      \let\leq=\leqslant 

    \fi
  \fi
\fi 

\ifCUPmtlplainloaded \else
  \ifAMStwofonts \else 
    \def\upi{\pi}
    \def\umu{\mu}
    \def\upartial{\partial}
  \fi
\fi

\title{On the dynamics of superfluid neutron star cores}

\author[N. Andersson and G. L. Comer]{N. Andersson${}^1$ and G. L. 
Comer${}^2$\\
${}^1$Department of Mathematics, University of Southampton, Southampton, 
UK \\
${}^2$Department of Physics, Saint Louis University, P.O. Box 56907, 
St. Louis, MO 63156-0907, USA}

\date{\today}

\begin{document}
	
\maketitle

\begin{abstract}
We discuss the nature of the various modes of pulsation of superfluid 
neutron stars using comparatively simple Newtonian models and the 
Cowling approximation.  The matter in these stars is described in terms 
of a two-fluid model, where one fluid is the neutron superfluid, which 
is believed to exist in the core and inner crust of mature neutron 
stars, and the other fluid represents a conglomerate of all other 
constituents (crust nuclei, protons, electrons, etc.).  In our model, 
we incorporate the non-dissipative interaction known as the  
entrainment effect, whereby the momentum of one constituent (e.g. the 
neutrons) carries along part of the mass of the other constituent.  We 
show that there is no independent set of pulsating g-modes in a 
non-rotating superfluid neutron star core, even though the linearized 
superfluid equations contain a well-defined (and real-valued) analog 
to the so-called Brunt-V\"ais\"al\"a frequency.  Instead, what we find 
are two sets of spheroidal perturbations whose nature is predominately 
acoustic. In addition, an analysis of the zero-frequency 
subspace (i.e. the space of time-independent perturbations) reveals 
two sets of degenerate spherioidal perturbations, which we interpret to be 
the missing g-modes, and two sets of toroidal perturbations.  
We anticipate that the degeneracy of all these zero-frequency modes
will be broken by the Coriolis force in the case of rotating stars.
To illustrate this we consider the toroidal pulsation modes of a slowly 
rotating superfluid star.  This analysis shows that the superfluid 
equations support a new class of r-modes in addition to those familiar 
from, for example, geophysical fluid dynamics.  Finally, the role of 
the entrainment effect on the superfluid mode frequencies is shown 
explicitly via solutions to dispersion relations that follow from a 
``local'' analysis of the linearized superfluid equations. 
\end{abstract}


\section[]{Introduction}

In the forty or so years since Migdal's \shortcite{M59} initial work, 
three important areas of research on superfluidity in neutron stars 
have emerged: (i) Nuclear physics studies of the pairing gap 
energies as functions of mass density (for recent reviews, see 
\cite{UL99,LS00}); (ii) observations and subsequent modeling 
of glitches (see, for instance, \cite{S89,L93} and references 
therein) and neutron star cooling \cite{T94,HH00}; and (iii) studies 
of the dynamics of superfluid neutron star cores in the Newtonian 
\cite{E88,ML91,M91,LM94,LM95,ulee,RP99,LM00} and general relativistic 
regimes \cite{C89,CL94,CL95-1,CL95-2,CL98,LSC98,CLL99,AC00}.  It is 
this third area that is the focus of the present discussion.  We 
want to understand how the pulsation properties of a neutron star 
changes when it cools below the temperature (a few times $10^{9}~{\rm 
K}$) at which the bulk of the interior is expected to become 
superfluid.  One of the underlying motivations for this work is the 
possibility that future gravitational-wave detectors may be sensitive 
enough to observe pulsating neutron stars, following (say) a glitch 
or a starquake \cite{astero1,astero2,astero3}.  An analysis of such 
observed data will provide a probe of the neutron star interior and 
should, at least in principle, allow us to infer the details of the 
supranuclear equation of state.  In particular, one would hope to be 
able to establish beyond any doubt that the core of a neutron star 
contains a superfluid and perhaps constrain some of the relevant 
parameters for (say) entrainment.  Obviously we can only hope to 
achieve this goal if the pulsation properties of a superfluid star 
differ significantly from those of an ``ordinary fluid'' neutron 
star.

The equations that describe ordinary (cold, inviscid etcetera)
fluid neutron stars consist of a single mass continuity equation 
and a single Euler equation that determines the fluid velocity.  But 
even this ``simple'' system yields an amazing diversity of modes of 
pulsation (see McDermott et al \shortcite{MVH88} for a fine 
discussion of many classes of modes), which includes the spheroidal 
f-, p-, and g-modes and the toroidal r-modes.  In reality cold 
neutron stars are significantly more complicated than what is 
implied by the ordinary fluid model.  The current understanding is 
that the outer regions of a mature neutron star consist of crust 
nuclei and an electron gas, which (beyond neutron drip) are 
everywhere permeated by superfluid neutrons. In the core, nuclei 
have dissolved leaving behind neutrons in a superfluid state, 
superconducting protons, and an ultra-relativistic gas of degenerate 
electrons.  In addition, more exotic particles (e.g. hyperons) may be 
present.  Deep in the core various condensates (of kaons, pions 
etcetera) may form, and deconfined quarks may also play a crucial 
role.  A priori, one might expect the pulsation spectrum of such a 
``real'' neutron star to be very complex, consisting of the various 
modes that exist for the ordinary fluid case, plus additional modes 
that arise because of new fluid degrees of freedom due to different 
particle species moving (more or less) independently of each other.  

The most striking ways that (pure) superfluids differ from 
ordinary fluids is that superfluids have zero viscosity and are 
locally irrotational \cite{TT86}.  The latter property, however, 
is compensated by the superfluid being threaded by an array of 
quantized vortices so that it can on macroscopic scales mimic the 
rotational behaviour of an ordinary fluid.  The neutron superfluid 
in a rotating neutron star is believed to contain such threading.  
The vortices also lead to an effective viscosity through the 
so-called entrainment effect and a consequence of it known as 
mutual friction.  In a mixture of the two superfluids Helium three 
and Helium four, it is known that a momentum induced in one of the 
constituents will cause some of the mass of the other to be carried 
along, or entrained \cite{AB75,TT86}.  The analog in neutron stars 
is an entrainment of some of the protons, say, by the neutrons 
\cite{ALS84,BJK96}.  Because of entrainment, the flow of neutrons 
around the neutron fluid vortices will also induce a flow in a 
fraction of the protons, leading to magnetic fields being formed 
around the vortices.  But since the electrons are coupled to the 
protons on very short timescales \cite{ALS84}, some will track 
closely the entrained protons.  Mutual friction is the dissipative 
scattering of these electrons off of the magnetic fields associated 
with the vortices.  

The comparatively simple model of neutron star superfluidity that 
will be used in this paper considers just two fluids, and the 
entrainment effect that acts between them.  One fluid consists of 
the superfluid neutrons that exist in the inner crust and core, 
whereas the other is a conglomerate of all the charged particles 
(i.e. the nuclei and electrons in the crust, and the superconducting 
protons and electrons in the core) that will be loosely referred to 
as ``protons.''  In comparison to the ordinary fluid case, the 
superfluid system of equations will consist of two mass continuity 
equations, and two Euler equations that determine the neutron and 
proton fluid velocities.  This model has been used in previous 
numerical studies of the linearized pulsations of superfluid 
neutron stars \cite{LM94,ulee,CLL99}.  However, it is important to 
point out that Mendell \shortcite{M91} has discussed the more 
general case where electromagnetic effects are explicitly included, 
and where the electrons (and even muons) are free to move 
independently of the protons, a net result of which is to delimit 
the dynamical timescales for which the simplified two-fluid model 
is appropriate. 

It is also relevant to mention that even before the equations 
were analyzed in detail, Epstein \shortcite{E88} (and then later 
Mendell \shortcite{M91}) argued, using a simple counting of the fluid 
degrees of freedom in a superfluid neutron star, that there should be 
a new class of modes that were later dubbed {\em superfluid} modes.  
They arise primarily because the neutrons and protons in the core are 
superfluid and no longer locked together in nuclei, thus leading 
to an increase in the number of fluid degrees of freedom.  It is also 
impressive that---using as an analogy a system of coupled 
pendulums---Mendell \shortcite{M91} 
argued that the distinguishing characteristic of 
the new modes should be the counter-motion of the neutrons with 
respect to the protons.  That is, in terms of a projection along the 
radial direction, the superfluid modes should have the protons moving 
oppositely to the neutrons, unlike an ordinary fluid mode that would 
have the neutrons and protons moving more or less in ``lock-step.''  
This picture has been confirmed by numerical (and simplified 
analytical) studies in both the Newtonian \cite{LM94,ulee} and general 
relativistic regimes \cite{CLL99}.  Another feature, shown by Lindblom 
and Mendell \shortcite{LM94}, is that the superfluid modes are driven 
by deviations from chemical equilibrium between the neutrons and 
protons.

At first sight, the existence of the superfluid modes appears to 
confirm one's intuition that the new fluid degrees of freedom would 
double the number of pulsation mode-families.  However, there is an 
open question in the literature regarding the g-modes in the 
superfluid case.  In particular, Lee's \shortcite{ulee} numerical 
results did not reveal {\em any} independent set of pulsating g-modes, 
much less a doubling.  This is perhaps strange, and certainly not 
consistent with the simple counting argument given above, since 
Reisenegger and Goldreich \shortcite{GR92} have shown convincingly 
that a composition gradient (for instance, a stable stratification 
in terms of the proton fraction) leads to the existence of propagating 
g-modes in non-rotating cold ordinary fluid neutron stars.  Even more 
puzzling is the fact that Lee showed that the ordinary fluid 
definition of the so-called Brunt-V\"ais\"al\"a frequency leads to 
real values, which is the usual indication that pulsating g-modes 
should exist.  His final conclusion was that the superfluid modes 
are to be found among the ordinary fluid f- and p-modes, and that 
there are no propagating g-modes in the cores of superfluid neutron 
stars, even if there exists a composition gradient.  Here, we 
will use a ``local'' analysis of the linearized superfluid equations 
to show that superfluid modes are predominately acoustic and to 
confirm Lee's numerical results that superfluidity prevents g-modes 
from pulsating.  

Of course, if there are no distinct g-modes then one is still left 
with the question of where they have gone. 
To find the answer, we 
will analyze the zero-frequency subspace of solutions to the 
superfluid equations of motion, following 
recent studies by Lockitch and Friedman \shortcite{lock} and 
Lockitch et al \shortcite{laf00} in the ordinary fluid case.  
This subspace is composed of the 
time-independent solutions to the linearized equations.  We find that 
the zero-frequency 
subspace appropriate to the superfluid equations is spanned by 
solutions that separate into two distinct classes: (i) Non-trivial 
local scalar matter (e.g. number densities, pressure, etc) 
perturbations accompanied by zero velocity perturbations, and (ii)  
degenerate velocity perturbations with vanishing local 
matter  perturbations. 
The first class of 
solutions simply takes static and spherically symmetric configurations 
and deforms them into other, nearby static and spherically symmetric 
configurations. The second class separates into 
independent sets of spheroidal and toroidal velocity perturbations. 
This is exacly analogous to the results 
for an ordinary fluid, but there is one important difference.
We find two sets of decoupled perturbations for each of the 
spheroidal/toroidal velocity perturbations. 

We interpret these results in the following way:   
Because the second class of solutions have vanishing local matter 
variations, it is natural to take the spheroidal 
solutions to be the missing 
g-modes. The existence of two independent sets of 
such solutions would then be in agreement with the intuitive 
notion that the 
fluid degrees of freedom are doubled in the superfluid case. 
The presence of spheroidal modes in the zero-frequency 
subspace is analogous to the ordinary fluid case when there is no 
stratification in the background star \cite{lock,laf00}.
The doubling of modes in the  
class of toroidal solutions is also unique to the superfluid.
Given these results, an analysis of rotating configurations 
becomes of prime importance. 
A comparison to  the ordinary fluid case 
\cite{lock,laf00}) suggests that rotation will break 
 the degeneracy of the various
zero-frequency and lead to 
{\em inertial modes} with true dynamics. 
Our results for the zero-frequency subspace suggests that, just like in a 
barotropic ordinary fluid star,  the 
typical inertial mode will be a hybrid mixture of spheroidal 
and toroidal velocity components in the non-rotating limit.

We take the first steps towards the study of inertial modes
in a superfluid star by considering the so-called r-modes 
(which correspond to purely toroidal velocities in the non-rotating 
limit). 
These modes are of particular relevance since it was recently discovered 
\cite{A98,FM98} that they are generically unstable to 
gravitational radiation via the so-called CFS 
(Chandrasekhar-Friedman-Schutz) mechanism \cite{C70,FS78,F78}.  In 
principle, the CFS mechanism can apply to any mode in a rotating 
neutron star, but viscous damping of various kinds tends to kill the 
instability before substantial amounts of gravitational radiation are 
emitted, except, apparently, for the r-modes 
\cite{LOM98,AKS99,owen,AK00}.  Remarkably, even though mutual friction 
in superfluid neutron stars is extremely effective at killing the CFS 
instability for f- and p-modes, Lindblom and Mendell \shortcite{LM00} 
have shown that it does not damp out the r-modes (except for a very 
small subset of the entrainment models employed).  But the main point 
for the present discussion is whether or not superfluidity leads to a 
new set of r-modes, thus making the problem richer than the normal 
fluid one.  Unfortunately, this is hard to discern from the discussion 
of Lindblom and Mendell \shortcite{LM00}.  While they argue 
 for the existence of some type of superfluid r-modes from their 
numerical results, in the sense that 
they see the counter-motion of the neutrons with respect to the 
protons, they find that these modes  differ little from the ordinary 
fluid r-modes at both the lowest- and second-order in the angular 
velocity of the background configuration.  Also, they argue that any 
analogy with the spheroidal type of superfluid modes discussed 
earlier is only superficial, since the fluid motion is not dominated 
by a deviation from chemical equilibrium.  We will argue below that a 
distinct set of superfluid r-modes does, in fact, exist.  Moreover, we 
will show that the frequencies of these modes, even at the lowest-order 
in angular velocity, differ from those of the ordinary fluid r-modes, 
and depend in an essential way on the strength of the entrainment 
effect. These results shed light on some of the (so far unexplained) 
peculiarities on the results regarding the effect of mutual friction 
on the unstable r-modes. 

Carter, Langlois and collaborators 
\cite{C89,CL95-1,CL95-2,CL98,LSC98,CL94} have been developing 
a system of equations that can describe superfluids in general 
relativistic neutron stars.  They are the relativistic analogs of 
the equations used by Lindblom and Mendell and Lee for the Newtonian 
regime.  Until recently, the relativistic equations had not been used
for modeling various scenarios but this situation is rapidly changing.
Comer, Langlois and Lin \shortcite{CLL99} have used the relativistic
equations to model the quasinormal modes of a simple neutron star 
model, and were successful at extracting the general relativistic 
version of the superfluid modes.  Andersson and Comer \shortcite{AC00} 
have built models of ``slowly rotating'' general relativistic 
superfluid neutron stars. A unique feature of their formalism is that 
the neutrons are not a priori forced to corotate with the protons.  
We will use the Newtonian limit of these equations for the main 
analysis presented below.  One reason for this is to show how to 
connect the general relativistic formalism with the Newtonian one.  
Another underlying motivation is to show how to incorporate the 
current microscopic models of entrainment in the general relativistic 
case. 

The layout of the paper is as follows: In Section~II, we review the 
standard reasoning that leads to a demonstration of p- and g-modes in 
cold (single fluid) neutron stars.  This will provide a convenient 
background for the analysis of the following three sections (III, IV, 
and V) of the perturbations of superfluid neutron stars.  Aside from the 
main points discussed above, a practical outcome of these calculations 
will be a simple, approximate formula for the superfluid mode frequencies 
that has not been given previously in the literature.  Largely just to 
simplify the analysis we use the so-called Cowling approximation, i.e.  
the variation in the gravitational field will be ignored. In Section~VI 
we unambiguously demonstrate the existence of two distinct classes of 
r-modes in a superfluid star.  A brief summary is given in Section~VII. 
Finally, in the Appendix we give the steps used to derive the Newtonian 
limit of the general relativistic superfluid field equations, and comment 
on the relation between our formulae and those used by Lindblom and 
Mendell.

\section[]{Stellar pulsation primer}

\subsection[]{The standard picture: p/g-modes}

Before turning to the superfluid case, it is useful to recall the
``standard'' theory of stellar pulsation \cite{cox,unno}.  For an 
ordinary fluid star, we need to consider the linearized version of
the perturbed Euler equation
\beq
    {\partial^2 \vec{\xi} \over \partial t^2} = {\delta \rho \over 
    \rho^2} \nabla P - {1 \over \rho} \nabla \delta P - \nabla \delta 
    \Phi
\eeq
(where $\vec{\xi}$ is the fluid displacement vector, $P$ denotes the 
pressure, $\rho$ is the density and $\Phi$ is the gravitational 
potential) as well as the (integrated form of) the continuity equation
\beq
    \delta \rho + \nabla \cdot (\rho \vec{\xi}) = 0 \ .
\eeq
We are using $\delta$ to denote Eulerian perturbations, while 
$\Delta$ will be used (in this section only) to indicate Lagrangian 
variations. 

It is customary to introduce the adiabatic index of the perturbations
$\Gamma_1$ via
\beq
    {\Delta P \over P} = \Gamma_1 {\Delta \rho \over \rho}
\eeq
or, in terms of the Eulerian variations, 
\begin{eqnarray}
    \delta P &=& {P \Gamma_1 \over \rho} \delta \rho + \vec{\xi} \cdot 
               \left[{P \Gamma_1 \over \rho} \nabla \rho - \nabla P 
               \right] \cr
              && \cr
             &\equiv& {P \Gamma_1 \over \rho} \delta \rho + 
               P \Gamma_1 (\vec{\xi}\cdot \vec{A})
\end{eqnarray}
which defines the Schwarzschild discriminant $\vec{A}$. 

For spherical stars we can now readily rewrite the Euler equation as
\beq
    {\partial^2 \vec{\xi} \over \partial t^2} = - \nabla 
    \left({\delta P \over \rho} \right) + {P \Gamma_1 \over \rho} 
    \vec{A} (\nabla \cdot \vec{\xi}) - \nabla \delta \Phi \ . 
    \label{eul1}
\eeq
Once the equation is written in this form we can deduce that the fluid 
motion is affected by (neglecting $\delta \Phi$) two restoring forces: 
the pressure variation and the buoyancy associated with $\vec{A}$.  
The latter is relevant whenever the star is stratified, either by 
entropy or compositional variations.  Since we are considering neutron 
stars we can to good approximation assume that the temperature is 
zero.  This means that we can neglect any internal entropy gradients.  
Still, as was pointed out by Reisenegger and Goldreich \shortcite{GR92}, 
we cannot assume that $\vec{A} = 0$ since any variation of the internal 
composition will lead to an effective buoyancy force acting on a fluid 
element.  For neutron stars, the contribution due to the varying 
proton fraction is likely to be the most important effect.

We want to study (\ref{eul1}) and try to infer the nature of the 
various modes of pulsation that the star may have.  In doing this we 
assume that the Cowling approximation is used, i.e. we neglect 
the variation $\delta \Phi$ in the gravitational potential.  If we 
further assume that the fluid element remains in {\em pressure
equilibrium} with its surroundings (in such a way that $\delta P = 0$) 
we have 
\beq 
    \Delta  P \approx \vec{\xi} \cdot \nabla P \equiv \rho \vec{\xi} 
     \cdot \vec{g} \ ,
\eeq
where we have used the standard definition of the local 
gravitational acceleration $\vec{g}$.  If we also use the continuity 
equation,
\beq
    \Delta \rho \equiv \delta \rho + \vec{\xi} \cdot \nabla \rho = 
    - \rho \nabla \cdot \vec{\xi} \label{delrho} \ , 
\eeq
we find that the radial component of the Euler equation becomes
\beq
    {\partial^2 \xi^r \over \partial t^2} = g \xi^r A \equiv - N^2 
    \xi^r 
\eeq
which defines the so-called Brunt-V\"ais\"al\"a frequency $N$.  In 
other words, we have oscillatory motion whenever $N^2 > 0$.  The 
resultant modes of pulsation are known as the g-modes, as they are 
essentially governed by gravity and internal stratification in the 
star.  Whenever $N^2<0$ the perturbation is unstable.

We can study the nature of the star's modes in more detail in the 
following way (cf. Reisenegger \& Goldreich 
\shortcite{GR92}).  Since $\vec{g}$ is purely radial 
for a spherical star the horizontal component of the Euler equation 
(\ref{eul1}) leads to
\beq
    \vec{\xi}_\perp = {1 \over \omega^2 \rho} \nabla_\perp 
                      {\delta P} \ ,
\eeq
where we have assumed that the perturbation has time-dependence 
$e^{i\omega t}$.  Combining this with (\ref{delrho}) we find that 
\begin{eqnarray}
    \Delta \rho &=& - {\rho \over r^2} {\partial \over \partial r} 
                   (r^2 \xi^r) - \rho \nabla_\perp \cdot \vec{\xi} 
                    \cr
                 && \cr
                &=& - {\rho \over r^2} {\partial \over \partial r} 
                  (r^2 \xi^r) + { l(l+1) \over \omega^2 r^2} 
                  \delta P \ ,
\end{eqnarray}
where we have assumed that the angular dependence of $\delta P$ can 
be represented by a single spherical harmonic $Y_{lm}(\theta,\varphi)$ 
(which is natural since the pressure variation is a scalar) such that
$$
  \nabla_\perp \cdot \nabla_\perp \delta P = - {l (l + 1) \over r^2} 
  \delta P \ .
$$
We also have the radial component of (\ref{eul1}):
\beq
    - \omega^2 \xi^r = - {\partial \over \partial r} \left(\delta 
                       P \over \rho \right) - {P \Gamma_1 A \Delta 
                       \rho \over \rho^2} \ . 
\eeq
In these two equations we can replace $\Delta \rho$ by noticing that
\beq
    \Delta \rho = {\rho \over P \Gamma_1} (\delta P -\rho g \xi^r)
\eeq
(having used the fact that $\vec{g} = - g \hat{e}_r$), and we get 
\beq
    {1 \over r^2} {\partial \over \partial r} (r^2 \xi^r) - {\rho g 
    \over P \Gamma_1} \xi^r = \left[{l (l + 1) \over \omega^2 r^2} - 
    {\rho \over P \Gamma_1}\right] {\delta P \over \rho} 
\eeq
and
\beq
    {1 \over \rho} {\partial \over \partial r} \delta P + {g \over P 
    \Gamma_1} \delta P = (\omega^2 + gA) \xi^r \ .
\eeq

Let us now introduce new variables
\begin{eqnarray}
      \hat{\xi}^r &=& {r^2 \xi^r \over \phi} \ , \\
      \delta \hat{P} &=& \phi \delta P \ ,
\end{eqnarray}
where
\beq
    \phi = \exp \left[\int g/c_s^2 d r\right] \label{phidef}
\eeq
and the sound speed is
\beq
    c_s^2 = {\Delta P \over \Delta \rho} = {P \Gamma_1 \over \rho} \ .
\eeq
With these definitions our two equations can be written
\beq
    {\partial \hat{\xi}^r \over \partial r} = \left[L_l^2 - \omega^2
    \right] {r^2 \delta \hat{P} \over \rho \omega^2 c_s^2 \phi^2} \ ,
\eeq
where we have introduced the so-called Lamb frequency
\beq
    L_l^2 = {l(l + 1) c_s^2 \over r^2} \ , 
\eeq
and 
\beq
    {\partial \delta \hat{P} \over \partial r} = \left[\omega^2 + g A 
    \right] {\rho \hat{\xi}^r \phi^2 \over r^2} \ .
\eeq

Given this we can reduce the problem to the following ordinary
differential equation for $\hat{\xi}^r$:
\begin{eqnarray}
    &&{d \over d r} \left\{{\rho \omega^2 c_s^2 \phi^2 \over r^2} 
    \left[L_l^2 - \omega^2 \right]^{-1} {d \hat{\xi}^r \over d r}
    \right\} - \cr
    && \cr
    &&[\omega^2 -N^2] {\rho \hat{\xi}^r \phi^2 \over r^2} 
    = 0 \ .
\end{eqnarray}
From this we can draw some very important conclusions.  We see that 
the problem reduces to the Sturm-Liouville form both for high and 
low frequencies.  For large $\omega^2$ we get
\beq
    {d \over d r} \left\{{\rho c_s^2 \phi^2 \over r^2} {d 
    \hat{\xi}^r \over d r} \right\} + [\omega^2 -N^2] {\rho 
    \hat{\xi}^r \phi^2 \over r^2} = 0 \ .
\eeq
Then standard Sturm-Liouville theory tells us that there will be an 
infinite set of modes which can be labelled by the number of radial 
nodes ($n$) of the various eigenfunctions, and for which $\omega_n \to 
\infty$ as $n \to \infty$.  In the opposite limit, when $\omega^2$ is 
small, the problem becomes
\beq
    {d \over d r} \left\{{\rho \phi^2 \over l (l + 1)} {d \hat{\xi}^r 
    \over d r} \right\} + \left[{N^2 \over \omega^2} - 1 \right] 
    {\rho \hat{\xi}_r \phi^2 \over r^2} = 0
\eeq
and we deduce that there will be another set of modes, with 
eigenfrequencies such that $\omega_n \to 0$ as $n \to \infty$.  

These sets of modes are the p- and g-modes, respectively, and we can 
estimate their frequencies in the following way: Assume that the 
perturbations have a characteristic wavelength $k^{- 1}$ (such that 
the various functions are proportional to $\exp(ikr)$).  Then we 
can readily deduce the simple dispersion relation
\beq
    k^2 = {1 \over c_s^2 \omega^2} (N^2 - \omega^2)(L_l^2 - \omega^2) 
          \ .
\eeq
Here we must have $\omega^2>0$ for stability, and we see that we will
have pulsating modes ($k^2>0$) in two different cases.  Either 
$\omega^2$ must be smaller than both $N^2$ and $L_l^2$, or it must be 
larger than both these quantities.  For $l >> k r$  we estimate the 
mode-frequencies as $\omega^2 \approx L_l^2$ for the p-modes and
$\omega^2 \approx N^2$ for the g-modes.

So far, we have discussed familiar textbook results \cite{cox,unno}. 
We did this in order to be able to compare and constrast these
results with the corresponding ones in the superfluid case. 
This is the aim of the rest of the paper.  

\subsection[]{The first step towards two fluids}

As a first step towards considering a superfluid star we approach the 
above pulsation problem in a somewhat indirect way.  Let us suppose 
that the star is composed of two distinct species of particles, which
we will think of as the superfluid neutrons in the core and the 
``protons'' as defined  in the Introduction.  We denote their 
respective number densities by $\n_\n$ and $\n_\p$.  Then standard 
thermodynamical considerations tell us that the pressure can be 
determined from
\beq
    d P = \n_\n d \mu_\n + \n_\p d \mu_\np \ ,
\eeq
where $\mu_i$ ($i = \n , \p$) are the two chemical potentials (which 
do not include the rest-masses).  In general, there should also be a 
term proportional to $d |\vec{v}_\n - \vec{v}_\p|^2$, where 
$\vec{v}_\n$ and $\vec{v}_\p$ are the neutron and proton velocities 
(cf. the Appendix), respectively, but since the background velocities 
are zero (or at least the same for both fluids) in the cases we will 
consider, such a term will not contribute in what follows.  By 
introducing
\beq
    \tilde{\mu}_n = {\mu_n \over m_n} \qquad , \quad
    \tilde{\mu}_p = {\mu_p \over m_p} \ ,
\eeq
we get the following relation between the Eulerian pressure variation 
and the variations in the two chemical potentials:
\beq
    \delta P = \rho_n \delta \tilde{\mu}_n  + \rho_p \delta 
               \tilde{\mu}_p \label{fund_rel} \ , \label{pressvar}
\eeq
where $\rho_\n = m_\n \n_\n$ and $\rho_\p= m_\p \n_\p$.

Chemical equilibrium for our system corresponds to $\mu_\n = \mu_\p$.  
It is natural to introduce a variable that describes the deviation 
from equilibrium introduced by the fluid motion. Thus we define 
$\delta \beta$ by 
\beq
    \delta \tilde{\mu}_\p = \delta \tilde{\mu}_\n + \delta \beta \ . 
                            \label{dbetadef}
\eeq
For simplicity, we have assumed that the two particle masses are 
equal: $m_\n = m_\p = m$. To define $\delta \beta$ in this particular 
way (in terms of the tilde variables) may seem peculiar, but it is 
useful since this variable then has exactly the same meaning as 
$\delta \beta$ in the series of papers by Lindblom and Mendell 
\shortcite{LM94,LM95}.  Anyway, the above relations enable us to write
\beq
    \delta \tilde{\mu}_\n = {\delta P \over \rho} - {\rho_\p \over 
                            \rho} \delta \beta \ .
\eeq
The corresponding thermodynamic condition is that
\beq
    d \tilde{\mu}_\n = {1 \over \rho} d P - {\rho_\p \over \rho} 
                       d \beta \ . 
\eeq
From this we can immediately read off that
\beq
    {1 \over \rho} = \left({\partial \tilde{\mu}_\n \over \partial P} 
                    \right)_{\beta} 
                    \quad , \quad
    {\rho_\p \over \rho} = - \left({\partial \tilde{\mu}_\n \over 
                             \partial \beta}\right)_P \ ,
\eeq
and using the equality of mixed partial derivatives, we can deduce the 
useful identity:
\beq
    \rho^2 {\partial \over \partial P} \left({\rho_\p \over \rho} 
          \right)_{\beta} = \left({\partial \rho \over \partial 
          \beta } \right)_P \ . \label{thermrel} 
\eeq

Let us now assume that the motion of each species of particles is 
determined by the variations in the chemical and gravitational 
potentials according to (see the Appendix for a justification)
\beq
    {\partial \delta \vec{v}_\n \over \partial t} + \nabla (\delta 
    \tilde{\mu}_\n + \delta \Phi) = 0 \label{euln}
\eeq
and
\beq
    {\partial \delta \vec{v}_\p \over \partial t} + \nabla (\delta 
    \tilde{\mu}_\p + \delta \Phi) = 0 \ .
\eeq
In view of the earlier relations, the second of the two equations can 
be written
\beq
    {\partial \delta \vec{v}_\p \over \partial t} + \nabla (\delta 
    \tilde{\mu}_\n + \delta \beta +\delta \Phi) = 0 \ . \label{eulp}
\eeq

Given (\ref{euln}) and (\ref{eulp}) we can make an important
observation: The neutrons and protons will only move together (in the 
sense that $\delta \vec{v}_\n = \delta \vec{v}_\p =\partial_t 
\vec{\xi}$) if $\delta \beta = 0$.  Intuitively this is obvious, since 
it simply says that the perturbation keeps the fluid in chemical 
equilibrium if the two species move together.  Yet, it will prove a 
useful observation later.  We note that if the two particle species 
move together we retain the standard Euler equation (\ref{eul1}) in the 
particular case $\vec{A} = 0$.  This is the first hint of a result that 
has repercussions on the pulsation properties of superfluid stars.  As 
we will now show, the case of barotropic perturbations ($\vec{A} = 0$) 
in the standard description, corresponds to $\delta \beta = 0$ in the 
two-fluid picture. 

We elucidate this correspondence in the following way: Introducing a 
new variable corresponding to the average velocity
\beq
    {\partial \vec{\xi}_+  \over \partial t} = {\rho_\n \over \rho} 
               \delta \vec{v}_\n + {\rho_\p \over \rho} \delta 
               \vec{v}_\p \ ,
\eeq
(i.e. $\rho~\partial_t \vec{\xi}_+$ is the total mass-density current)
and combining (\ref{euln}) and (\ref{eulp}), we readily get
\beq
    {\partial^2 \vec{\xi}_+ \over \partial t^2} + \nabla \delta 
    \tilde{\mu}_\n + \nabla \delta \Phi + {\rho_p \over \rho} \nabla 
    \delta \beta = 0 
\eeq
or
\beq
    {\partial^2 \vec{\xi}_+ \over \partial t^2} + \nabla \left(
    {\delta P \over \rho}\right) + \nabla \delta \Phi - {\partial 
    \over \partial P} \left({\rho_\p \over \rho} \right) \delta \beta 
    \nabla P = 0 \ . \label{eulc}
\eeq
Comparing (\ref{eulc}) to the standard Euler equation (\ref{eul1}), 
and identifying $\vec{\xi}_+ = \vec{\xi}$, we see that we must have
\beq
    {\partial \over \partial P} \left({\rho_p \over \rho}
    \right)_{\beta} \delta \beta \nabla P =  {P \Gamma_1 \over \rho} 
    \vec{A} (\nabla \cdot \vec{\xi}_+) \ . \label{sdisc}
\eeq
In other words, the one-fluid Schwarzschild discriminant is intimately 
linked to the variation $\delta \beta$ in the two-fluid picture, and 
indeed $\delta \beta = 0$ implies $\vec{A} = 0$. This is quite natural
since (in the case of a chemical composition gradient) the Schwarzschild 
discriminant describes the extent to which a given perturbation drives 
a fluid element away from chemical equilibrium.  This observation 
implies that the pulsation properties of a two-fluid model with $\delta 
\beta = 0$ should be analogous to the case of barotropic perturbations 
for which there are no non-trivial g-modes.  In other words, in this 
case one expects only to find a set of p-modes.

\section[]{The equations governing a perturbed superfluid}

We now consider the full two-fluid model for superfluid 
neutron stars, i.e. consider the superfluid neutrons as being 
coupled to a conglomerate of charged particles (protons, electrons, 
nuclei etcetera).  Adopting the notation of the discussion in the 
previous section we first rewrite the linearized form of the equations 
(\ref{fullconsvn})-(\ref{fulleulerp}) from the Appendix in terms of a 
set of variables that are intimately linked to the physics of the 
system.  The first two of these are the average velocity $\partial_t 
\vec{\xi}_+$ and the deviation from beta-equilibrium $\delta \beta$ 
that we introduced in Section~2.  Together with these we use the 
pressure variation $\delta P$ and a variable reflecting the relative 
motion of the neutrons and the protons.  In order to retain complete 
correspondence with the equations used by Lindblom and 
Mendell\footnote{This is obviously not necessary but we want 
to avoid unnecessary confusion among readers who are
well acquainted with the relevant literature.} we define
\beq
    {\partial \vec{\xi}_- \over \partial t} = {\rho_\n \rho_\p \over \det 
               \rho}~(\delta \vec{v}_\p - \delta \vec{v}_\n) \ ,
\eeq
where we note that the case $\vec{\xi}_- = 0$ corresponds to the 
two species of particles moving together.  We have introduced  
\begin{eqnarray}
    \det \rho &=& \rho_{\n \n}\rho_{\p \p} - \rho_{\n \p}^2 \ , \cr
               && \cr
    \rho_\n &=& \rho_{\n \n} + \rho_{\n \p} \ , \cr
               && \cr
    \rho_\p &=& \rho_{\p \p} + \rho_{\n \p} \ . \label{detrho}
\end{eqnarray}
The mass-density matrix element $\rho_{\n\p}$ is what allows for the 
entrainment effect \cite{AB75}, and it can be rewritten as
\beq
    \rho_{\n \p} = - \epsilon \rho_\n \ ,
\eeq  
where $\epsilon$ is the so-called entrainment parameter.  We also have
\beq
\rho_{nn} = \rho_n(1+\epsilon) \ , \quad \rho_{pp} = \rho_p + \epsilon 
\rho_n \ .
\eeq
In the model used by Lindblom and Mendell \shortcite{LM00} one has
\beq
    \epsilon = {\rho_\p \over \rho_\n} \left({m_\p \over m^*_\p} - 1
               \right) \ , \label{entrain}
\eeq
where $m^*_\p$ is the effective proton mass (which enters 
because the protons form a Fermi liquid and it is thus their 
associated quasiparticle features that are paramount, see 
Sj\"oberg \shortcite{S76}).

The first continuity equation (i.e. the sum of the linearized versions 
of (\ref{fullconsvn}) and (\ref{fullconsvp}) of the Appendix) takes 
the form
\beq
    \delta \rho + \nabla \cdot (\rho \vec{\xi}_+) = 0 \label{cont2a}
\eeq
and we also arrive (by adding the linearized versions of 
(\ref{fulleulern}) and (\ref{fulleulerp}) of the Appendix) at the same 
equation of motion as in Section~II:
\beq
    {\partial^2 \vec{\xi}_+ \over \partial t^2} + \nabla \left({\delta 
    P \over \rho} \right) - {1 \over \rho^2} \left({\partial \rho 
    \over \partial \beta} \right)_P \delta \beta \nabla P = 0 \ .
    \label{eul2a}
\eeq
We have also assumed (for reasons of clarity) that the Cowling 
approximation is made, i.e. $\delta \Phi = 0$.

In addition (by subtracting the linearized versions of 
(\ref{fulleulern}) and (\ref{fulleulerp}) of the Appendix) we have a 
second equation of motion relating $\vec{\xi}_-$ and $\delta \beta$:
\beq
    {\partial^2 \vec{\xi}_ - \over \partial t^2} + \nabla \delta \beta 
                = 0 \label{eul2b}
\eeq
and (after some manipulations\footnote{The derivation of this 
equation is somewhat involved.  The first step is to ``invert'' the 
velocities to find
\begin{eqnarray}
    \delta \vec{v}_\n &=& {\partial \over \partial t} \left(\vec{\xi}_+ - 
                          {\det \rho \over \rho \rho_\n} \vec{\xi}_-
                          \right) \ , \nonumber \cr 
                       && \cr 
    \delta \vec{v}_\p &=& {\partial \over \partial t} \left(\vec{\xi}_+ + 
                          {\det \rho \over \rho \rho_\p} \vec{\xi}_-\right)
                          \ . \nonumber
\end{eqnarray}
Next one inserts the relationships obtained from (\ref{densityvars}) into 
the linearized form of the proton mass-continuity equation, using also 
the inverted velocities given above.  Finally, one uses the 
linearized form of the neutron mass-continuity equation to obtain an 
equation for $\partial_i \xi^i_+$.}) the final continuity equation can 
be written 
\begin{eqnarray}
    &&\rho^2 {\partial \over \partial P} \left( {\rho_p \over \rho} 
    \right)_\beta {\delta P \over \rho} + {\rho_\n^2 \over \rho} 
    {\partial \over \partial \beta} \left({\rho_\p \over \rho_\n}
    \right)_P \delta \beta + \cr
    && \cr
    &&\rho \vec{\xi}_+ \cdot \nabla 
    \left({\rho_\p \over \rho} \right) + \nabla \cdot \left({\det 
    \rho \over \rho} \vec{\xi}_-\right) = 0 \ . \label{cont2b} 
\end{eqnarray}
These four equations are identical to those derived and used by
Lindblom and Mendell \shortcite{LM94,LM95}. 

In these equations it is, however, difficult to discern the meaning, 
and thus importance, of the various thermodynamic derivatives that 
appear.  Fortunately, the situation can be clarified somewhat by 
introducing the two local sound speeds that our two-fluid system of 
equations admit (just like their mathematical twins, the superfluid 
equations constructed for superfluid helium by Landau \cite{P}).  
Using the standard technique of expanding the fluid variables as 
plane waves, and considering the background fluid to be homogeneous 
and at rest, we find that these sound speeds are obtained as 
solutions for $u^2$ from the following quadratic: 
\begin{eqnarray}
    &&0 = \left[u^2 - c^2_\n\right] \left[u^2 - {m_{\p} \over m^*_{\p}} 
        c^2_\p\right] + {\rho_{\p} \over \rho_{\n}} \left[\left(
        {m_{\p} \over m^*_{\p}} - 1\right) \right. \times \cr
       && \cr
       &&\left.\left(\left[2 c^2_{\n\p} - 
        c^2_\n\right] u^2 + c^2_\n c^2_\p - \left[1 + {\rho_{\p} \over 
        \rho_{\n}}\right] c^4_{\n\p}\right) - c^4_{\n\p}\right] \ , 
\end{eqnarray}
where
\begin{eqnarray}
    c^2_\n &\equiv& \rho_{\n} {\partial \tilde{\mu}_{\n} \over 
                \partial \rho_{\n}} \ , \cr
           && \cr 
    c^2_\p &\equiv& \rho_{\p} {\partial \tilde{\mu}_{\p} \over 
                \partial \rho_{\p}} \ , \cr
           && \cr 
    c^2_{\n\p} &\equiv& \rho_{\n} {\partial \tilde{\mu}_{\n} \over 
                \partial \rho_{\p}} = \rho_{\n} {\partial 
                \tilde{\mu}_{\p} \over \partial \rho_{\n}} \ .
\end{eqnarray}
When the ratio $\rho_{\p}/\rho_{\n}$ is small we see that the 
sound speeds are essentially $c^2_\n$ and $(m_{\p}/m^*_{\p}) c^2_\p$.  
In terms of these definitions we can now write the various 
thermodynamic derivatives in the (still exact) forms
\begin{eqnarray}
    {1 \over c^2_{\rm eq}} &\equiv&
    \left({\partial \rho \over \partial P} \right)_\beta = {x_{\n} 
    \over c^2_\n} \left[1 - {x_{\p} \over x_{\n}} \left({c_{\n\p} 
    \over c_\n}\right)^2 \left({c_{\n\p} \over c_\p}\right)^2
    \right]^{- 1} \times \cr
    && \cr
    &&\left[1 + {x_{\p} \over x_{\n}} \left({c_\n 
    \over c_\p}\right)^2 - 2 {x_{\p} \over x_{\n}} \left({c_{\n\p} 
    \over c_\p}\right)^2\right] \ , \label{eqspeed} 
\end{eqnarray}
\begin{eqnarray}
    \left({\partial \rho \over \partial \beta} \right)_P &=& \rho_{\n} 
     {x_{\p} \over c^2_\p} \left[1 - {x_{\p} \over x_{\n}} 
    \left({c_{\n\p} \over c_\n}\right)^2 \left({c_{\n\p} \over 
    c_\p}\right)^2\right]^{- 1} \times \cr 
    && \cr
    &&\left[1 - \left({c_\p \over c_\n}\right)^2 - \left(1 - {x_{\p} 
    \over x_{\n}}\right) \left({c_{\n\p} \over c_\n}\right)^2\right] \ , 
    \label{drhodbeta}
\end{eqnarray}
\begin{eqnarray}
    {\partial \over \partial \beta} \left({ \rho_p \over \rho_n} 
    \right)_P &=& {x_{\p} \over c^2_\p} \left[1 - {x_{\p} \over 
    x_{\n}} \left({c_{\n\p} \over c_\n}\right)^2 \left({c_{\n\p} 
    \over c_\p}\right)^2\right]^{- 1} \times \cr
    && \cr
    &&\left[1 + {x_{\p} \over x_{\n}} \left({c_\p \over c_\n}\right)^2 
    + 2 {x_{\p} \over x_{\n}} \left({c_{\n\p} \over c_\n}\right)^2
    \right] \ , \label{drhoprhon}
\end{eqnarray}
and
\beq
    {1 \over \Delta} = {c^2_\n c^2_\p \over \rho^2 x_{\n} 
                        x_{\p}} \left[1 - {x_{\p} \over x_{\n}} 
                        \left({c_{\n\p} \over c_\n}\right)^2 
                        \left({c_{\n\p} \over c_\p}\right)^2\right]
                        \ , \label{Delta}
\eeq
where $x_{\n,\p} = \rho_{\n,\p}/\rho$ and we have used 
(\ref{LMTD1})--(\ref{LMTD4}) given in the Appendix.  (Also, one should 
not confuse the use of the $\Delta$ symbol here with its earlier role 
as a Lagrangian variation.)  The utility of these expressions is that 
they will illuminate the role of the proton fraction in determining 
the order of magnitude contributions from individual terms in the 
dispersion relations that will be written below.

An identity that follows from (\ref{eqspeed})--(\ref{Delta}) above, and 
which will be used later, is
\beq
    c_{\rm eq}^2 \left({\partial \rho \over \partial \beta}
      \right)_P^2 - \rho_n^2 {\partial \over \partial \beta} \left(
      {\rho_\p \over \rho_\n}\right)_P = - c^2_{\rm eq} \Delta \ .
\eeq
In anticipation of later results, and in order to facilitate 
comparisons between the ordinary fluid and superfluid cases, we define 
for the superfluid system of equations two frequencies:
\beq
    {\cal L}_l^2 \equiv {l (l + 1) c_{\rm eq}^2 \over r^2} 
\eeq
and
\beq
    {\cal N}^2 \equiv {g^2 \over c^2_{\rm eq} \Delta}~\left({\partial 
               \rho \over \partial \beta} \right)^2_P \ .
\eeq
The first of these corresponds to the standard Lamb frequency, while the 
second has (as will be discussed later)  similar character to the
Brunt-V\"ais\"al\"a frequency.

\section[]{Oscillations of nonrotating superfluid stars}

We now want to analyze the superfluid perturbation equations in a 
way similar to that used in Section~II for the standard pulsation 
problem.  There are several reasons for doing this.  The most obvious 
one being that we want to contrast the pulsation properties of a 
superfluid star with the standard ordinary fluid results in order to 
see whether future observations may be able to distinguish between the 
two cases.  From the general nature of the equations one might expect 
that the character of the various modes of oscillation may be rather 
different in the two cases.  After all, in the superfluid case we are 
allowing the two fluids to move more or less independently and so 
would seem to have brought in additional fluid dynamical degrees of 
freedom.   On the other hand, the g-modes in the standard case depend 
crucially on the stable stratification mainly associated with the 
varying chemical composition \cite{GR92}.  It is thus not at all clear 
what will happen to these modes if we allow the neutrons and the 
protons to move relative to one another.

\subsection[]{Two simple limiting cases}

Just as in the ordinary fluid case it is interesting, and potentially 
instructive, to consider what happens if we freeze various
degrees of freedom.  For example, the equations governing 
$\vec{\xi}_+$ and $\delta P$ reduce to the standard one-fluid 
equations (cf. Section~II) and it seems reasonable to think that these 
variables should therefore reflect the ordinary fluid properties.  At 
the same time, we have seen that $\delta \beta$ was to a certain extent 
accounted for in the standard picture via the buoyancy and the 
Schwarzschild discriminant $A$.  Finally, the variable $\vec{\xi}_-$ 
is clearly unique to the two-fluid system and could therefore be 
expected to bring some new features to the pulsation problem.

Let us first consider the case when the neutrons and protons move in 
such a way that they remain in chemical equilibrium.  This would 
correspond to $\delta \beta = 0$, and we immediately see that this 
requires $\vec{\xi}_- = 0$ as well.  In other words, the neutrons and 
the protons must move together (quite intuitively).  From the analysis 
in Section~II (cf. the discussion that leads to (\ref{sdisc})) we 
already know that the two equations (\ref{cont2a}) and (\ref{eul2a}) 
reduce to the standard equations for barotropic (nonstratified) 
stars ($A = 0$).  Hence, the only non-trivial modes we expect 
to find are the p-modes.  In this case one can easily show that they 
will have frequency
\beq
    \omega^2_o \approx {\cal L}_l^2
\eeq
noting, however, that $c_s^2$ is being replaced by $c_{\rm eq}^2$, 
i.e. the sound speed defined in (\ref{eqspeed}).  This is, of course, 
exactly the ordinary fluid result of Section~II in the limit $A 
\to 0$.

As a side remark it is interesting to note that one would not normally 
expect the above assumptions to lead to the two remaining equations, 
(\ref{eul2b}) and (\ref{cont2b}), also being satisfied.  However, in 
the present case (\ref{eul2b}) is trivial and, for particular models, 
one can also satisfy (\ref{cont2b}). This requires 
\beq
    {\partial \over \partial P} \left({\rho_p \over \rho}
    \right)_\beta = 0 \ .
\eeq  
In other words, we need to have $\rho_\p \propto \rho$ or $\rho_\p = 
0$.  The latter does, of course, correspond to a star in which there 
is only one species of particles so the equations of motion must 
reduce to the standard ones for a single fluid.

In Section~II we saw that the nature of the g-modes could be deduced
by requiring that the motion be such that the fluid remains in 
pressure equilibrium, $\delta P = 0$.  Let us consider the equations
from the previous section in this case.  Conveniently ``forgetting'' 
(for the moment) the conservation of mass equation (\ref{cont2a}) we 
need to consider
\beq
    {d \delta \beta \over d r} - {\omega^2 \rho \over r^2 \det \rho} 
              Z = 0 \ ,
\eeq
and
\begin{eqnarray}
    {d Z \over d r} &=&  
    \left[\left({\det \rho \over c^4_{eq} \Delta} {\cal L}^2_l + 
    {\cal N}^2\right) {1 \over \omega^2} - \left(1 + \left[{c_{\rm eq} 
    \over g} {\cal N}\right]^2\right)\right] \times \cr
    && \cr
    &&{r^2 c^2_{\rm eq} \Delta \over \rho}~\delta \beta \ ,
\end{eqnarray}
which follows from (\ref{eul2a})--(\ref{cont2b}), 
once we separate the radial and horizontal components of the equations 
(as in Section~II) and introduce the new variable ($\xi^r_-$ is the 
radial component of $\vec{\xi}_-$)
\beq
    Z = {\det \rho \over \rho} r^2 \xi^r_- \ ,
\eeq
as well as use the various definitions and relations given at the end of 
Section~III.  We now combine the two equations to find
\begin{eqnarray}
    &&{d \over d r} \left\{{\rho \over r^2 c^2_{\rm eq} 
    \Delta} \left[\left(1 + \left[{c_{\rm eq} \over g} {\cal N}
    \right]^2\right) - \right.\right. \cr
    && \cr
    &&\left.\left.\left({\det \rho \over c^4_{\rm eq} \Delta} 
    {\cal L}^2_l + {\cal N}^2\right) {1 \over \omega^2}\right]^{-1} 
    {d Z \over d r}\right\} + {\omega^2 \rho \over r^2 \det 
    \rho} Z = 0
\end{eqnarray} 
and it is easy to see that, just as in Section~II, we have a 
Sturm-Liouville problem for large $\omega^2$.  Hence, we expect there 
to exist a set of modes for which $\omega_\n^2 \to \infty$ for the 
large overtones (with index $n \to \infty$). 

To elucidate the nature of these modes, let us derive the appropriate 
dispersion relation.  Assuming that the perturbation variables behave 
as $\exp(ikr)$ we readily arrive at
\beq
    k^2 = {c^2_{\rm eq} \Delta \over \det \rho} \left[\left(1 + \left[
          {c_{\rm eq} \over g} {\cal N}\right]^2\right) \omega^2 - 
          \left({\det \rho \over c^4_{\rm eq} \Delta} {\cal L}^2_l + 
          {\cal N}^2\right)\right]
\eeq
and in the limit when $l >> k r$ we find
\beq
    \omega^2_s \approx \left(1 +  \left[{c_{\rm eq} \over g} {\cal N}
                       \right]^2\right)^{- 1} \left({\det \rho \over 
                       c^4_{\rm eq} \Delta} {\cal L}^2_l + {\cal N}^2
                       \right) \ .
\eeq
Assuming that the sound speeds are well-behaved if $x_\p$ is small, 
then we can infer from (\ref{eqspeed}) that $c^2_{\rm eq}$, and hence 
${\cal L}^2_l$, are roughly independent of the proton fraction to 
leading order.  We also infer from (\ref{detrho}) and (\ref{Delta}) 
that the combination $\det \rho/\Delta$ is independent of the proton 
fraction to leading order.  Finally, we see from (\ref{Delta}) and 
(\ref{drhodbeta}) that  
\beq
    {1 \over \Delta}~\left({\partial \rho \over \partial \beta}
                     \right)^2_P \sim x_p \ ,
\eeq 
which implies that 
\beq
    {\cal N}^2 \sim x_\p \ .
\eeq

Thus, to linear order in the proton fraction we see
\beq
     \omega^2_s \approx {\cal L}^2_l \left(1 -  \left[{c_{\rm eq} 
                        \over g} {\cal N}\right]^2\right) {\det \rho 
                        \over c^4_{\rm eq} \Delta} + {\cal N}^2 \ .
\eeq
To the lowest-order in the proton fraction, the dominant contribution 
to $\omega_s$ is
\beq
    \omega^2_s \approx {\cal L}^2_l {\det \rho \over c^4_{\rm eq} 
    \Delta} \approx {m_\p \over m^*_p}~{l (l + 1) \over r^2}~c^2_\p 
    \ .
\eeq
Hence we deduce that these ``superfluid'' pulsation modes are 
essentially governed by the proton sound speed $c^2_\p$, 
and also that the entrainment effect plays an important role in 
determining their frequencies.  We also notice that contributions due 
to  ${\cal N}$ are 
negligible at this order.  This approximate result provides a natural 
explanation for the result of Comer et al \shortcite{CLL99} that the 
ordinary and superfluid frequencies are virtually the same when the 
adiabatic index of the neutrons (treated as a relativistic polytrope) 
is set equal to the index for the protons (also treated as a 
relativistic polytrope).  In that case, the sound speeds of the two 
fluids are virtually the same. Furthermore, in the models considered in 
Comer et al \shortcite{CLL99} 
 there is no entrainment since $m^*_\p$ is set equal to $m_\p$ .

\subsection[]{The fully coupled case}

Having established the existence of two distinct classes of pulsation 
modes in a superfluid star we now return to the full problem, and 
consider the modes of pulsation for the system of equations 
(\ref{cont2a})-(\ref{cont2b}). 

Proceeding in the now familiar way, we separate the radial and 
horizontal parts of (\ref{eul2a}) and (\ref{eul2b}), and then combine 
the results with the two conservation laws (\ref{cont2a}) and 
({\ref{cont2b}).  This way we arrive at a set of four equations
\begin{eqnarray}
    &&{1 \over r^2} {d\over d r} \left(r^2 \xi^r_+ \right)
    - {g \over c_{\rm eq}^2} \xi^r_+ + \left[ {1 \over c_{\rm eq}^2} - 
    {l (l + 1) \over \omega^2 r^2} \right] {\delta P \over \rho} + \cr
    && \cr
    &&\qquad {1 \over \rho} \left( {\partial \rho \over \partial \beta} 
    \right)_P \delta \beta = 0 \ , \cr
    && \cr
    &&{d \over dr} \delta P + {g \over c_{\rm eq}^2} 
    \delta P - \rho \omega^2 \xi^r_+ + g \left({\partial \rho \over 
    \partial \beta} \right)_P \delta \beta = 0 \ , \cr
    && \cr
    &&Z - {r^2 \over \omega^2}  {\det \rho \over \rho} {d\over 
    dr} \delta \beta = 0 \ , \cr
    && \cr
    &&{1 \over r^2} {dZ \over dr} + \left[{\rho_\n^2 
    \over \rho}  {\partial \over \partial \beta} \left({\rho_\p \over 
    \rho_n} \right) - {l (l + 1) \over \omega^2 r^2 } {\det \rho \over 
    \rho}\right] \delta \beta + \cr
    && \cr
    &&\qquad \left({\partial \rho \over \partial \beta} \right)_P 
    \left[{\delta P \over \rho} - g \xi^r_+\right] = 0 \ ,
\end{eqnarray} 
where $\xi^r_+$ is the radial component of $\vec{\xi}_+$ and the 
variable $Z$ was defined in  Section~IVA.

As in Section~II the first two equations simplify considerably if we 
introduce the integrating factor $\phi$ as defined by (\ref{phidef}) 
(replacing $c_s$ with $c_{\rm eq}$, of course), and then work with the 
variables $\delta \hat{P}= \phi \delta P$ and $\hat{Y} = r^2 \xi^r_+ /
\phi$.  If we also introduce the characteristic wavelength of the 
pulsation through $\exp(ikr)$, our first two equations take the form
\beq
    i k \hat{Y} + \left[{r^2 \over c_{\rm eq}^2} - {l (l + 1) \over 
    \omega^2}\right] {\delta \hat{P} \over \rho \phi^2} = - {r^2 \over 
    \rho} \left({\partial \rho \over \partial \beta} \right)_P {\delta 
    \beta \over \phi}  
\eeq
and
\beq
    i k \delta \hat{P} - {\rho \omega^2 \over r^2} \phi^2 \hat{Y} = 
    - g \left({\partial \rho \over \partial \beta} \right)_P \phi 
    \delta \beta \ . 
\eeq
These two relations can easily be combined to provide a relation 
between $\delta \hat{P} $ and $\delta \beta$:
\begin{eqnarray}
    \left[{l (l + 1) + k^2 r^2 \over r^2} c_{\rm eq}^2 - \omega^2 
    \right] \delta \hat{P} &=& [i k g + \omega^2]~\times \cr
    && \cr
    &&c_{\rm eq}^2 \phi \left({\partial \rho \over \partial \beta} 
      \right)_P \delta \beta \ .
\end{eqnarray}
After some similar manipulations, one can show that the two remaining 
perturbation equations lead to
\begin{eqnarray}
    &&[\omega^2 - i k g] \left({\partial \rho \over \partial \beta} 
    \right)_P \delta \hat{P} = \left[{l(l + 1) + k^2 r^2 \over r^2} 
    \det \rho - \right. \cr
    && \cr
    &&\qquad \left.\omega^2 \rho_\n^2 {\partial \over \partial \beta} 
    \left( {\rho_p \over \rho_\n} \right) + g^2 \left({\partial \rho 
    \over \partial \beta} \right)_P^2\right] \phi \delta \beta \ .
\end{eqnarray}  
These two equations can be combined to provide the dispersion relation 
for waves propagating in a superfluid star
\begin{eqnarray}
    &&[\omega^4 + k^2 g^2] c_{\rm eq}^2 \left({\partial \rho \over 
    \partial \beta}\right)_P^2 = \left[{l(l + 1) + k^2 r^2 \over 
    r^2} c_{\rm eq}^2 - \omega^2 \right] \times \cr
    && \cr 
    &&\qquad \left[{l(l + 1) + k^2 r^2 \over r^2} \det \rho - \omega^2 
    \rho_\n^2 {\partial \over \partial \beta} \left({\rho_\p \over 
    \rho_\n}\right)_P + \right. \cr
    && \cr
    &&\qquad \left.g^2 \left({\partial \rho \over \partial \beta} 
    \right)_P^2 \right] \ . \label{maindisp}
\end{eqnarray}
In principle, this equation contains information equivalent to the 
dispersion relation of Lindblom and Mendell \shortcite{LM94} (cf. 
their equation (82)).

We can do one final rewriting of (\ref{maindisp}) as a quadratic in 
$\omega^2$, which we could then solve.  Using the various definitions 
and relations listed at the end of Section~III, the fully coupled 
dispersion relation takes the final form
\begin{eqnarray}
    && 0 = \omega^4 - \left\{\left[1 + {k^2 r^2 \over l (l + 1)}\right] 
          \left[1 + {\det \rho \over c^4_{\rm eq} \Delta}\right] 
          {\cal L}^2_l  + \right. \cr
       && \cr
       && \qquad \left. \left[1 + \left(1 + {k^2 r^2 \over l (l + 1)}
          \right) \left({c_{\rm eq} \over g} {\cal L}_l\right)^2
          \right] {\cal N}^2\right\} \omega^2 + \cr 
       && \qquad {\cal L}^2_l \left\{\left[1 + {k^2 r^2 \over l (l + 1)}
          \right]^2{\det \rho \over c^4_{\rm eq} \Delta} {\cal L}^2_l 
          + {\cal N}^2\right\} \ . \label{quadratic}
\end{eqnarray}
It is, however, not particularly instructive to write down the formal 
solution to this equation.  It is much better to first simplify it 
somewhat. This provides a better insight into the relevant physics of 
the solution.

\subsection[]{Results for a small proton fraction}

Our aim now is to use the results of the previous section to infer 
how the presence of a superfluid affects the pulsation modes of a 
nonrotating neutron star core.  In principle, we have drawn the main 
conclusions already.  There will exist two more or less distinct 
classes of modes.  One of these comprise the standard p-modes, while 
the other has unique properties (although see comments below) due to 
the presence of the superfluid.  Both these families of modes are 
such that the eigenfrequencies $\omega_n^2\to \infty$ as the 
mode-number $n \to \infty$.  In other words, the two sets of modes 
will be interlaced in the pulsation spectrum of the star.  These are, 
however, only qualitative results and it would clearly be useful to 
make more quantitative estimates.  

The key point to simplifying the fully coupled dispersion relation is 
that ${\cal N}^2 \sim x_\p$ which means that the frequencies can easily 
be determined to linear order in the proton fraction.  Doing this (and 
considering also the limit where $l >> k r$) yields the two solutions
\beq
    \omega^2_{o} \approx {\cal L}^2_l \left(1 + \left[1 - {\det \rho 
    \over c^4_{\rm eq} \Delta}\right]^{- 1} \left[{c_{\rm eq} \over g} 
    {\cal N}\right]^2\right)
\eeq
and
\beq
    \omega^2_{s} \approx {\det \rho \over c^4_{\rm eq} \Delta} 
    {\cal L}^2_l \left(1 - \left[1 - {\det \rho \over c^4_{\rm eq}
    \Delta}\right]^{- 1} \left[{c_{\rm eq} \over g} {\cal N}\right]^2
    \right) + {\cal N}^2 \ . \label{sfreq}
\eeq
To lowest-order in the proton fraction, we recover the ordinary, i.e. 
$\omega^2_o$, and superfluid, i.e. $\omega^2_s$, solutions 
respectively.  

Let us now address one of the main questions that motivated the present 
investigation: What happens to the g-modes when the neutron star core 
becomes superfluid?  We recall the discussion by Reisenegger and 
Goldreich \shortcite{GR92} that showed that a non-superfluid neutron star 
will have a distinct set of g-modes whose nature is determined by the 
chemical composition gradient associated with the varying proton 
fraction.  From Section~II we know that these modes will have 
eigenfrequencies such that $\omega_n^2\to 0$ as $n \to \infty$.  We have 
already shown that there will not be a family of modes with this character 
in the superfluid case.  Thus, even though we clearly have a varying 
proton fraction, a superfluid neutron star core will have no pulsating 
g-modes, in accordance with the numerical results of Lee 
\shortcite{ulee}.  This is part of the answer to the question posed  
above, and in order to provide the rest of the answer, we will need to 
consider the space of 
time-independent solutions to the linearized equations, i.e. the 
zero-frequency subspace.  But before doing 
that, we want to emphasize some consequences of having a varying proton 
fraction.  

The effect of a composition gradient will be analyzed as follows: In 
the ordinary fluid case one can estimate the Schwarzschild discriminant 
due to a variation in the proton fraction as (see Unno et al 
\shortcite{unno} and the discussion in Lee \shortcite{ulee} that 
surrounds Lee's equation (47))
\beq
    A = {1 \over \rho} \left({\partial \rho \over \partial 
        \rho_\p}\right)_P \hat{e}_r \cdot \nabla \rho_p
\eeq
(assuming the star has zero temperature).  This then leads to g-mode
frequencies
\beq
    \omega^2_g \approx N^2 = - g A = - {g \over \rho} \left({\partial 
    \rho \over \partial \rho_\p}\right)_P \hat{e}_r \cdot \nabla 
    \rho_p \ , \label{gfreq} 
\eeq
or
\beq
    \omega^2_g \approx N^2 \approx g^2 \left({\partial \rho \over 
               \partial \rho_\p}\right)_P \left({\partial \rho_\p 
               \over \partial P}\right)_{\beta} \approx x_\p 
               \left({g \over c_\p}\right)^2 \ .
\eeq
Now consider the dominant contribution to the superfluid mode frequency 
(\ref{sfreq}) when $x_\p$ and $c^2_\p$ are both small; it is simply
\beq
    \omega^2_s \approx {\cal N}^2 \approx x_\p \left({g \over c_\p}
                       \right)^2 \ . 
\eeq
From this we see that $\omega^2_g \approx N^2 \approx {\cal N}^2 
\approx \omega^2_s$ when the proton fraction and $c^2_\p$ are small.
Thus we conclude that, even though a superfluid neutron star core 
does not support a distinct class of pulsating g-modes, the buoyancy 
associated with internal composition gradients can be relevant. Of course, 
the effect depends crucially on the proton fraction and cannot be 
distinguished unless $c^2_\p$ is small.  

\section{The zero-frequency subspace}

We will now focus our attention on the perturbations of the superfluid 
equations that belong to the zero-frequency subspace. That is, after 
perturbing (and linearizing) the equations, we assume that the velocity 
perturbations, as well as $\delta \n_\n$ and $\delta \n_\p$, are 
time-independent (cf. \cite{lock,laf00}) .  
We then get from (\ref{fullconsvn}) and 
(\ref{fullconsvp}):
\beq
    0 = \partial_i \left(\rho_\n \delta v^i_\n\right) 
        \quad , \quad
    0 = \partial_i \left(\rho_\p \delta v^i_\p\right) \ . \label{veleqn}
\eeq
Meanwhile, the two linearized Euler equations and the equation 
for the linearized Newtonian potential lead to
\beq
    0 = \partial_i \left(\delta \Phi + \delta \tilde{\mu}_\n \right) 
        \quad , \quad 
    0 = \partial_i \left(\delta \Phi + \delta \tilde{\mu}_\p \right) \ ,
        \label{euleqn}
\eeq
and
\beq
    \nabla^2 \delta \Phi = 4 \pi G \left(\delta \rho_\n + \delta 
                          \rho_\p\right) = 4 \pi G \delta \rho\ .
                          \label{poispert}
\eeq
From these equations we immediately deduce that we can split the 
perturbations into two independent
sets.  The first set has the perturbed velocities 
vanishing but the other perturbations non-zero.  The second set has 
$\delta v^i_\n$ and $ \delta v^i_\p$ nonvanishing while the local matter
perturbations $\delta\rho_n = \delta \rho_p = \delta \tilde{\mu}_n = 
\delta \tilde{\mu}_p = \delta \Phi = 0$.  The general solution will be a 
superposition of solutions from each set. 

\subsection{Neighbouring spherical equilibria}

We begin by interpreting the meaning of the first set of perturbations 
distinguished above.  By taking the difference between the two equations 
in (\ref{euleqn}) it is easy to show that we must have
\beq
    \partial_i \delta \beta = 0 \ .
\eeq
Given this, we can also show that the sum of the two equations leads to 
\beq
    \partial_i \delta P = - \delta \rho \partial_i \Phi - \rho 
                            \partial_i \delta \Phi \ . 
\eeq
These two equations follow simply from the perturbed versions of the 
equations that determine the unperturbed configuration, i.e.
\beq
    \partial_i \left(\Phi + \tilde{\mu}_\n\right) = 0 \quad , \quad
    \partial_i \left(\Phi + \tilde{\mu}_\p\right) = 0 \ ,
\eeq
using also the earlier equation that relates pressure variations to 
chemical potential variations.  And just as for the equilibrium equations 
we can show that the solutions must have spherical symmetry.  Thus, the 
interpretation of this set of perturbations becomes clear: They simply 
correspond to a neighbouring spherical equilibrium model.  Provided that 
$\delta \beta = 0$ the perturbed model is in chemical equilibrium.

\subsection{Non-trivial velocity perturbations}

We now turn to the other set of perturbations, for which the
three-velocities are non-zero and represent  
convective currents.
  We first have to account for the fact 
that perturbations of a spherical star can be decoupled into two 
different classes: spheroidal and toroidal.  

Spheroidal velocity perturbations take the form
\begin{equation}
  \delta \vec{v}_j =  { 1 \over r}  \left( W_j(r) , V_j(r) 
  \partial_\theta , { V_j(r)
\over \sin \theta} \partial_\varphi \right) Y_{lm} \ , \qquad j=n,p
\end{equation}
where we have used a standard decomposition in terms of spherical 
harmonics.  With these definitions we readily get from (\ref{veleqn})
\begin{eqnarray}
    0 &=& l (l + 1) \n_n V_\n - {d \over d r} \left(r \n_n W_\n\right) 
          \ , \label{gone}\\
     && \cr
    0 &=& l (l + 1) \n_p V_\p - {d \over d r} \left(r \n_p W_\p\right) 
          \label{gtwo}\ .
\end{eqnarray}
Foregoing here a detailed discussion of boundary conditions, we can 
establish that non-trivial solutions exist.  In particular, we see 
that there are two functions ($W_\n$ and $W_\p$, say) that can be 
freely specified.  Because they are spheroidal, and have vanishing 
pressure 
and chemical potential perturbations, we identify these solutions as the 
g-modes ``missing'' from the time-dependent perturbations.  In this 
respect, 
we see that superfluid g-modes behave like those in the barotropic
ordinary fluid case, i.e.  when 
the equation of state of the perturbations is the same as that of the 
background \cite{lock,laf00}.  However, the doubling of the modes 
(because there are two free functions) is unique to the superfluid.

Toroidal three-velocity perturbations can be written as 
\beq
  \delta \vec{v}_j = { 1 \over r} \left( 0, {U_j(r) \over \sin \theta} 
                     \partial_\varphi , 
                  - U_j(r) \partial_\theta \right)  Y_{lm}  
                  \ , \qquad j=n,p
\eeq           
For velocity perturbations of this form Eq. (\ref{veleqn}) is 
automatically satisfied for arbitrary $U_\n$ and $U_\p$.  Since there 
are two functions that can be freely specified we deduce that there will 
be two sets of toroidal modes in the zero-frequency subspace.  When the 
star is rotating this should lead to the presence of two classes of 
r-modes. We will verify this in the next section. 

\section{Rotating superfluid stars: Qualitative insights into the 
r-modes}

In the last two years various aspects of rotating neutron stars have 
been under intense scrutiny following the discovery \cite{A98,FM98} 
that the emission of gravitational waves drives the so-called r-modes 
unstable.  Initial studies of the problem indicated that the r-mode 
instability might cause a newly born neutron star to spin down to a 
rotation rate comparable to that inferred from observation for the 
Crab pulsar \cite{LOM98,AKS99}, in the process radiating gravitational 
waves that may well prove detectable with the generation of large-scale 
interferometers due to come online in the next few years \cite{owen}.  
One issue of utmost importance for the r-mode problem concerns the 
role of superfluidity \cite{AK00}.  So for example was it originally 
thought that dissipation due to the superfluid mutual friction would 
counteract the instability in a significant way.  The only available 
calculation of this effect, due to Lindblom and Mendell 
\shortcite{LM00}, suggests a rather different picture, however.  Their
 results indicate that the mutual friction will typically not be able 
to suppress the r-mode instability.  But the results also suggest that 
the effect becomes dominant for certain values of the entrainment 
parameter $\epsilon$.  Clearly, this problem is far from well 
understood at the present time. 

Our aim in this section is to study the r-modes at a qualitative level
(comparable to our study of spheroidal p- and g-modes in the previous 
sections).  This would seem a natural starting point for a discussion 
of the r-modes of superfluid stars, and as we will see it provides 
insights that may well explain various features observed in the 
numerical work by Lindblom and Mendell \shortcite{LM00}.

To analyze the r-mode problem we will extend Saio's \shortcite{S82} 
vorticity argument to the two-fluid problem.  Thus we focus on the 
two Euler equations.  When expressed in terms of the variables 
$\vec{\xi}_+$ and $\vec{\xi}_-$, these equations can be written
\begin{eqnarray}
    &&\partial^2_t \vec{\xi}_+ + (\vec{u} \cdot \nabla) \partial_t 
    \vec{\xi}_+ + \partial_t \vec{\xi}_+ \cdot \nabla \vec{u} + 
    \nabla \delta \Phi + \cr
    && \cr
    &&\qquad \nabla \delta \tilde{\mu}_\n + x_\p \nabla 
    \delta \beta = 0
\end{eqnarray}
and
\begin{eqnarray}
    &&\partial^2_t \vec{\xi}_- + (\vec{u} \cdot \nabla) \partial_t 
    \vec{\xi}_- - \partial_t \vec{\xi}_-\cdot \nabla \vec{u} + \nabla 
    \delta \beta + \cr
    && \cr
    && \qquad 2 {\det \rho \over \rho_\n \rho_\p} (\partial_t 
    \vec{\xi}_- \cdot\nabla) \vec{u} = 0 \ .
\end{eqnarray}
Here we have assumed that the protons and neutrons corotate in the 
unperturbed case.  The relevant rotation velocity is denoted by 
$\vec{u}$ in the above equations.  We note that it may be desirable 
to relax the assumption of corotation of the two background fluids 
in order to model a real neutron star (cf. the discussion in Andersson 
and Comer \shortcite{AC00}). The resultant problem is, however, much 
more complicated than the present one and we will return to it in 
future investigations. 

We first translate these equations into the rotating frame. Then we get
\beq
    \partial^2_t \vec{\xi}_+ + 2 \vec{\Omega} \times \partial_t 
    \vec{\xi}_+ + \nabla \delta \Phi + \nabla \delta \tilde{\mu}_\n + 
    x_\p \nabla \delta \beta = 0 \label{roteul1} 
\eeq
and
\begin{eqnarray}
    &&\partial^2_t \vec{\xi}_- + 2 \vec{\Omega} \times \partial_t 
    \vec{\xi}_- + \nabla \delta \beta - 2 \left[1 - {\det \rho \over 
    \rho_\n \rho_\p}\right] \times \cr
    && \cr
    &&\qquad (\partial_t \vec{\xi}_- \cdot \nabla) \vec{u} = 0 \ . 
    \label{roteul2}
\end{eqnarray}

The next step involves assuming that the mode is horizontal to leading 
order, taking the curl of the above two equations and focusing on the 
radial component of the resultant relations.  Consider first 
(\ref{roteul1}), which readily yields
\beq
    \partial_t \left[\nabla \times \partial_t \vec{\xi}_+ + 2 \nabla 
    \times \vec{\Omega} \times \vec{\xi}_+\right]_r = 0
\eeq
since 
\beq
    \nabla \times \left(x_\p \nabla \delta \beta \right) = \nabla x_\p 
    \times \nabla \delta \beta 
\eeq
has a vanishing radial component for a slowly rotating (still spherical)
star.  We now use
\begin{eqnarray}
    \left[\nabla \times \vec{\Omega} \times \vec{\xi}_+\right]_r &=&
    \left\{\vec{\Omega} (\nabla \cdot  \vec{\xi}_+) - \vec{\xi}_+ 
    (\nabla \cdot \vec{\Omega}) + \right. \cr
    && \cr
    &&\left. (\vec{\xi}_+ \cdot \nabla) 
    \vec{\Omega} - (\vec{\Omega} \cdot \nabla) \vec{\xi}_+\right\}_r 
    \cr
    && \cr 
    &\approx& \left[(\vec{\xi}_+ \cdot \nabla) \vec{\Omega}\right]_r 
    \ .
\end{eqnarray}
After these manipulations we have arrived at
\beq
    \partial_t \left[\partial_t (\nabla \times \vec{\xi}_+) + 2 
    (\vec{\xi}_+ \cdot \nabla) \vec{\Omega}\right]_r \approx 0 \ . 
\eeq
Finally, we insert in this equation the assumption that the mode we 
are interested in is purely toroidal, i.e. is of the form
.
\beq
    \vec{\xi}_+ = r \left(0 , {T \over \sin \theta} \partial_\varphi , 
                  - T \partial_\theta \right) e^{i\omega_r t} Y_{lm} 
                  \ , \label{torus}
\eeq
Given this assumption we immediately find that 
these modes must have frequency
\beq
    \omega_r = {2 m \Omega \over l (l + 1)} \ ,
\eeq
in the rotating frame.  This is, of course, the standard r-mode 
result \cite{AK00}. 

Let us now go through the same exercise for the second Euler equation, 
(\ref{roteul2}), that appears in the superfluid case. The radial 
component of the curl of this equation can be written (in the rotating 
frame)
\begin{eqnarray}
    &&\partial_t \left\{(\nabla \times \partial_t \vec{\xi}_-) + 2 
    \nabla \times (\vec{\Omega} \times \vec{\xi}_-) - \right. \cr
    && \cr
    &&\qquad \left. 2 \nabla \times \left[\left(1 - { \det \rho \over 
    \rho_n \rho_p}\right) (\vec{\xi}_- \cdot \nabla) \vec{u}\right]
    \right\}_r = 0 \ . 
\end{eqnarray}
Here the first two terms are identical to those we encountered in the 
analysis of (\ref{roteul1}), but the last term is new.  This term can 
be written
\beq
    2 \nabla \times \left(\gamma \vec{\xi}_- \cdot \vec{u}\right) = 2 
    \left(\nabla \gamma\right) \times (\vec{\Omega} \times 
    \vec{\xi}_-) + 2 \gamma \nabla \times (\vec{\Omega} \times 
    \vec{\xi}_-) \ , \label{relat}
\eeq
where we have introduced
\beq
    \gamma =  1 - {\det \rho \over \rho_\n \rho_\p} \ . 
\eeq
It is clear that the first term of the right-hand side of 
(\ref{relat}) does not contribute to the radial component.  The second 
term, however, does have a non-vanishing radial component.  We readily 
find that
\begin{eqnarray}
    &&[2 \nabla \times \left(\gamma \vec{\xi}_- \cdot \vec{u} \right)]_r 
    = {2 \gamma \Omega \over r \sin \theta} \times \cr
    && \cr
    &&\qquad \left[\partial_\theta (\sin \theta \cos \theta \xi^\theta_-
    ) + \partial_\varphi (\cos \theta \xi^\varphi_-)\right]
\end{eqnarray}
and if we assume that the vector $\vec{\xi}_-$ is toroidal, i.e. takes 
the form (\ref{torus}), we have
\begin{eqnarray}
    [2 \nabla \times \left(\gamma \vec{\xi}_- \cdot \vec{u} \right)]_r 
    &=& {2 \gamma \Omega T \over \sin \theta} \times \cr
     && \cr
    &&\left[\partial_\theta (\cos \theta \partial_\varphi Y_{lm} ) - 
    \partial_\varphi (\cos \theta \partial_\theta Y_{lm})\right] \cr
     && \cr
    &=& - 2 i m \Omega \gamma T Y_{lm} \ .
\end{eqnarray}
We can now combine this result with the final result obtained from 
(\ref{roteul1}) to deduce that we will have modes with frequency
\beq
    \omega_r = {2 m \Omega \over l (l + 1)} {\det \rho \over \rho_\n 
               \rho_\p} \ .
\eeq
From the definition of $\det \rho$ it follows that this is identical to
\beq
    \omega_r = {2 m \Omega \over l(l + 1)} \left[1 + \epsilon\left(1 + 
               {\rho_\n \over \rho_\p}\right)\right] \ ,
\label{sfrm}\eeq
where $\epsilon$ is the entrainment coefficient. This is a very 
interesting result
since it demonstrates the existence of a distinct class of superfluid 
r-modes, the properties of which are to a large extent determined by 
the entrainment parameter. 

Based on the above analysis we can now discuss the general nature of 
the toroidal modes of rotating superfluid neutron star cores (even 
though we should advice some caution since (\ref{sfrm}) is 
$r$-dependent and therefore strictly speaking only describes an r-mode 
in a thin shell).  We have seen that, just like in the non-rotating 
case, there will be two distinct classes of modes.  The first 
corresponds to fluid motion such that the neutrons and the protons 
flow together, and the resultant modes are analogous to the standard 
perfect fluid r-modes.  These are the modes that the calculations of 
Lindblom and Mendell \shortcite{LM00} were aimed at studying.  For 
the second class, the neutrons and the protons are counter-moving.  
This class of r-modes has not previously been discussed in the 
literature (although see the discussion below).  It is interesting to 
note that the two classes of modes are (to leading order) degenerate 
in the absence of entrainment.  This is, of course, rather intuitive 
since the two fluids are then effectively uncoupled and the Coriolis 
force affects each fluid separately.  

We note that the existence of two distinct classes of r-modes in 
superfluid stars provides a likely explanation for the somewhat 
peculiar results obtained by Lindblom and Mendell \shortcite{LM00} 
regarding the effect of mutual friction on the r-mode instability.  
In their study Lindblom and Mendell found that the mutual friction 
dissipation timescale was sufficiently long that this effect would 
not suppress the unstable r-modes for most values of $\epsilon$.  
But they also discovered that mutual friction would become dominant 
for certain values of $\epsilon$, cf. their figure~6.  Furthermore, 
they noted that while the variable $\delta \beta$ was small in most 
cases, it became large for the particular values of $\epsilon$ at 
which mutual friction was found to be important.  Given our present 
conclusions regarding the r-modes in superfluid stars, we can 
interpret the Lindblom-Mendell result in the following way: There are 
two distinct classes of r-modes in a superfluid star.  One of these 
classes, the one for which the neutrons and protons flow together, 
is typically not rapidly damped by mutual friction.  The other class 
of modes, however, is such that the two fluids are counter-moving.  
Since mutual friction tends to damp relative motion between the two 
fluids these modes will be strongly affected by mutual friction.  
This then makes the Lindblom-Mendell results quite natural.  For 
certain values of the entrainment parameter they have simply found 
the second class of modes rather than the first. 

The situation may, however, be yet more complicated.  It is likely 
relevant to make the following observation: In many situations where 
the mode-spectrum is studied for a family of stellar models for 
which a single parameter is varied (such as the crustal shear modulus) 
one finds that the modes undergo what are known as ``avoided 
crossings.''  So for example have recent studies of neutron stars 
with an elastic crust shown that the r-modes associated with the core 
fluid show avoided crossings with the toroidal shear modes in the 
crust for certain values of the rotation rate.  Avoided crossings 
between pulsation modes are, in fact, a regular occurence in studies 
of more complicated stellar models \cite{cox,unno}.  It is interesting 
to note that, well away from the crossing point, the different modes 
have a  distinct nature, but as one approaches the avoided crossings 
the two modes become similar in nature and when one reaches ``the 
other side'' they have completely exchanged properties (the core 
r-mode has become a crust shear mode and vice versa).  In view of 
our qualitative analysis and the numerical results of Lindblom and 
Mendell, it would seem natural to predict that a more detailed 
investigation of the superfluid r-mode problem will unveil analogous 
avoided crossings between the co- and counter-moving r-modes as the 
entrainment coefficient is varied. 

\section[]{Conclusions}

In this paper we have compared and contrasted the pulsation properties 
of a superfluid neutron star core (represented by a relatively simple 
two-fluid model) with the familiar (textbook) results for a normal 
fluid star.  This study provides a qualitative understanding of the 
nature of the oscillation modes of a superfluid neutron star, and thus 
provides a theoretical fundament for various numerical results in this 
area \cite{LM94,ulee}. We have shown that a non-rotating 
superfluid neutron star 
core exhibits two distinct families of pulsation modes.  Our 
approximate results provide a deep insight into the physics of these 
modes, with one set of modes corresponding to the two fluids moving 
in ``lock-step'' while the second  family (the modes that are unique to a 
superfluid) correspond to the two fluids being counter-moving.  
We have shown that both sets of modes are crucially governed by the 
acoustic properties of the two fluids.  This is in clear contrast to 
the normal fluid case, where the two families of modes, the p- and 
g-modes, are acoustic and governed by buoyancy (mainly due to chemical 
composition gradients), respectively.  We have investigated what 
happens to the g-modes as the star becomes superfluid and have 
confirmed Lee's numerical results that there are no 
propagating g-modes in a superfluid core.
In addition, we have shown that the ``missing'' g-modes 
can be found in the zero-frequency subspace.  Finally, we have 
shown how the various modes are affected by the parameters of 
entrainment. 

Basically, our results are important for two reasons. Firstly, 
they provide a clear and concise description of the problem and
the general nature of pulsating superfluid stars. Thus they 
fill what we perceive as a gap in the existing literature.
In particular, we believe we have resolved some open questions 
regarding g-modes in the superfluid case.  Secondly, 
the results indicate that future observations of neutron star 
mode oscillations, eg. by a highly sensitive generation of 
gravitational wave detectors, could potentially help constrain
the parameters of the large scale superfluidity that is believed to
exist in the core of mature neutron stars. We will discuss this
exciting possibility elsewhere.     

In addition to studying non-rotating stellar cores, we have presented
a qualitative analysis of the r-modes in a slowly rotating 
superfluid core. We have demonstrated that there will be two distinct
families of such modes, and argued that the interplay between 
these modes (eg. via so-called avoided crossings as the entrainment
parameter is varied) may provide an understanding of puzzling 
numerical results obtained by Lindblom and Mendell \shortcite{LM00}.
Our current analysis clearly shows that the r-mode problem for 
superfluid stars is likely to be much richer than has previously 
been appreciated.  In particular since, in addition to the r-modes, 
we anticipate that there will be a large class of ``hybrid'' inertial 
modes 
\cite{lock,laf00}.  No attempts to study such modes in a superfluid 
star have yet been made.  A better understanding of these various 
issues is clearly needed in view of the fact that the r-modes have 
been shown to be unstable due to the emission of gravitational waves, 
and that this instability may govern the spin-evolution of neutron 
stars during various phases of their lives. We therefore plan to 
investigate this problem in greater detail in the near future.

\section*{Acknowledgements}

We wish to thank J. Friedman, K. D. Kokkotas, D. Langlois, and R. Prix 
for fruitful discussions.  GLC gratefully acknowledges partial support 
from a Saint Louis University SLU2000 Faculty Research Leave Award as 
well as EPSRC in the UK via grant number GR/R52169/01, and the warm 
hospitality of the Center for Gravitation and Cosmology of the 
University of Wisconsin-Milwaukee and the University of Southampton 
where part of this research was carried out.  NA acknowledges generous 
support from PPARC via grant numberPPA/G/S/1998/00606, the EU Network 
``Sources of Gravitational Radiation,'' as well as the EU programme 
``Improving the Human Research Potential and the Socio-Economic
Knowledge Base'' (research training network contract HPRN-CT-2000-00137).

\appendix

\section[]{Newtonian limit of the general relativistic 
superfluid equations}

The purpose of this Appendix is to derive the Newtonian version of 
the general relativistic superfluid field equations.  There are 
several motivations for doing this.  For conceptual reasons we want 
to be able to compare our relativistic calculations \cite{CLL99,ACL00} 
to previous results in this research area (in particular those of 
Lindblom and Mendell \shortcite{LM94,LM95,LM00}) most of which were 
obtained in the Newtonian regime.  We also feel that it is useful to 
address various problems on a qualitative level before working out the 
detailed answers.  Given that Newtonian calculations are often simpler, 
it is natural to address the Newtonian problem first. In addition to 
this, we hope that a study of the Newtonian limit will provide 
insight into ways of extending various Newtonian entrainment models
to the relativistic regime. 

\subsection[]{The general relativistic formalism and Newtonian limit}

We begin by recounting the formalism that has been used to model 
general relativistic superfluid neutron stars 
\cite{CL95-1,CL95-2,LSC98,CLL99,AC00}. The fluid dynamical degrees of 
freedom are described by $\n^{\mu}_\n$, the conserved neutron number 
density current, and $\n^{\mu}_\p$, the conserved proton number density 
current.  The fundamental scalar in the formalism is the so-called 
``master'' function $\Lambda$, which is a function of the three 
scalars $\n_\n^2 = - \n^\n_{\rho} \n^{\rho}_\n$, $\n_\p^2 = - 
\n^\p_{\rho} \n^{\rho}_\p$, and $x^2 = - \n^\p_{\rho} \n^{\rho}_\n$, 
where $\n^\n_\mu = g_{\mu \nu} \n^\nu_\n$ and $\n^\p_\mu = g_{\mu \nu} 
\n^\nu_\p$.  The quantity $- \Lambda$ corresponds to the total 
thermodynamic energy density of the entire fluid.

There are two momenta $\mu_{\mu}$ and $\chi_{\mu}$ which are 
dynamically, and thermodynamically, conjugate to $\n^{\mu}_\n$ and 
$\n^{\mu}_\p$.  They are defined by a general variation (that keeps 
the spacetime metric fixed) of $\Lambda(\n^2_\n,\n^2_\p,x^2)$, i.e. 
\beq
     \d \Lambda = \mu_\rho \d \n^\rho_\n + \chi_\rho \d \n^\rho_\p \ ,
\eeq
where 
\beq
      \mu_{\mu} = \K^{\n\n} \n^\n_{\mu} + \K^{\n\p} \n^\p_{\mu} 
                 \quad , \quad
     \chi_{\mu} = \K^{\p\p} \n^\p_{\mu} + \K^{\n\p} \n^\n_{\mu} 
                  \ , \label{muchidef}
\eeq
and
\beq
    \K^{\n\p} = - {\partial \Lambda \over \partial x^2} 
               \quad , \quad 
    \K^{\n\n} = - 2 {\partial \Lambda \over \partial \n^2_\n} 
               \quad , \quad
    \K^{\p\p} = - 2 {\partial \Lambda \over \partial \n^2_\p} 
               \ . \label{coef1}
\eeq
In this general relativistic context the entrainment effect is seen   
in that the momentum $\mu_{\mu}$, say, is a linear combination of 
$\n^{\mu}_\n$ and $\n^{\mu}_\p$.  The generalized pressure $\Psi$ is 
given by  
\beq
     \Psi = \Lambda - \n^{\rho}_\n \mu_{\rho} - \n^{\rho}_\p 
            \chi_{\rho} \ .
\eeq
The equations of motion consist of two conservation equations,
\beq
    \nabla_{\mu} \n^{\mu}_\n = 0 
              \qquad , \qquad 
    \nabla_{\mu} \n^{\mu}_\p = 0 \ , \label{coneq}
\eeq
and two Euler type equations, which can be conveniently written in 
the compact form 
\beq
     \n^{\mu}_\n \nabla_{[\mu} \mu_{\nu]} = 0 
            \qquad , \qquad 
     \n^{\mu}_\p \nabla_{[\mu} \chi_{\nu]} = 0 \ , \label{eueqn} 
\eeq
where the square braces `[~]' indicate antisymmetrization on the 
indices.  

In order to derive the Newtonian limit, the general relativistic 
field equations will be written to order $c^0$ where $c$ is the 
speed of light.  Formally, the Newtonian equations will then be 
obtained in the limit that the speed of light $c$ becomes infinite.  
The gravitational potential, denoted $\Phi$, is assumed to be small 
in the sense that
\beq
     - 1 << {\Phi \over c^2} \leq 0 \ .
\eeq
To order $c^0$ the metric can be written as
\beq
    ds^2 = - c^2 \left(1 + {2 \Phi \over c^2}\right)  d t^2 + 
           \d_{i j} d x^i  d x^j \ ,
\eeq
where the $x^i$ ($i = 1,2,3$) are Cartesian-like coordinates.  

In the superfluid field equations there are two four-velocities that 
must be considered.  This means that we must take into account the 
fact that the fluids define two different proper times: one for the 
neutrons, to be denoted $\tau_{\n}$, and the second for the protons, 
to be denoted $\tau_{\p}$.  The two different fluid trajectories are 
then obtained from the functions
\begin{eqnarray}
  x_{\n}^{\mu}(\tau_{\n}) &=& (t(\tau_{\n}),x_{\n}^i(\tau_{\n}))
        \ , \cr
        && \cr
  x_{\p}^{\mu}(\tau_{\p}) &=& (t(\tau_{\p}),x_{\p}^i(\tau_{\p})) \ .
\end{eqnarray}
The respective four-velocities of the neutron and proton fluids are 
thus given by
\beq
    u^{\mu}_\n = {dx_{\n}^\mu \over d\tau_{\n}} \quad , 
              \quad 
    u^{\mu}_\p = {dx_{\p}^\mu \over d\tau_{\p}} \ .
\eeq
Because of the choice of coordinates, the four-velocities satisfy 
$u^\n_{\mu} u^{\mu}_\n = - c^2$ and $u^\p_{\mu} u^{\mu}_\p = - c^2$, 
where $u^\n_\mu = g_{\mu \nu} u^\nu_\n$ and $u^\p_\mu = g_{\mu \nu} 
u^\nu_\p$.  

We will turn this around and use a global time $t(\tau_{\n}) = 
t(\tau_{\p}) \equiv t$ as the parameter for both curves.  In this 
case, the two proper times will be given by
\begin{eqnarray}
   - d s_\n^2 &=& c^2 d\tau_{\n}^2 = c^2 \left(1 + {2 \Phi \over c^2} 
                  - {\d_{ij} v_{\n}^i v_{\n}^j \over c^2}\right) dt^2 
                  \ , \cr
                       && \cr
  - d s_\p^2 &=& c^2 d\tau_{\p}^2 = c^2 \left(1 + {2 \Phi \over c^2} - 
                   {\d_{ij} v_{\p}^i v_{\p}^j \over c^2}\right) dt^2 \ ,
\end{eqnarray}
where $v_{\n}^i = dx_{\n}^i/dt$ and $v_{\p}^i = d x_{\p}^i/dt$ are the 
Newtonian three-velocities of the neutron and proton fluids, 
respectively.  Each three-velocity is considered to be small in the 
sense that
\beq
     {\left|v_{\n}^i\right| \over c} << 1 \quad , \quad 
     {\left|v_{\p}^i\right| \over c} << 1 \ .
\eeq
Hence, to the correct order the four-velocity components are given by
\beq
u^t_\n = 1 - {\Phi \over c^2} + {v_{\n}^2 \over 2 c^2} 
      \ , \ 
u^i_\n = v_{\n}^i \ ,
\eeq
and
\beq
u^t_\p = 1 - {\Phi \over c^2} + {v_{\p}^2 \over 2 c^2} 
      \ , \ 
u^i_\p = v_{\p}^i \ ,
\eeq
where $v_{\n}^2 = \d_{ij}v_{\n}^i v_{\n}^j$ and $v_{\p}^2 = \d_{ij}
v_{\p}^i v_{\p}^j$. 

Note that the two particle number currents are now written as
\beq
    \n^{\mu}_\n = \n_\n \left(u^{\mu}_\n/c\right) 
                  \quad , \quad
    \n^{\mu}_\p = \n_\p \left(u^{\mu}_\p/c\right) \ .
\eeq
To the correct order, one finds for $x^2 = - \n_{\mu}^\n \n^{\mu}_\p$ 
that
\beq
    x^2 = \n_\n \n_\p \left(1 + {w^2 \over 2 c^2}\right) \ ,
\eeq
where 
\beq
    w^2 = \d_{ij} \left(v_{\n}^i - v_{\p}^i\right) 
                  \left(v_{\n}^j - v_{\p}^j\right) \ .
\eeq

In order to write the Euler equations with the terms to the required 
order, it is necessary to explicitly break up the ``master'' function 
into its mass part and internal energy part $E$, i.e., to write 
$\Lambda$ as
\beq
   \Lambda = - \left(m_{\n} \n_\n + m_{\p} \n_\p\right) c^2 - 
               E(\n^2_\n,\n^2_\p,x^2) \ .
\eeq
In this context, $E$ is small in the sense that, for instance,
\beq
     0 \leq {E \over m_{\n} \n_\n c^2} << 1 \ . 
\eeq
It is also convenient to use a different choice for the independent 
variables that more closely agrees with what is used for Newtonian 
superfluids, which is the triplet of variables $(\n^2_\n,\n^2_\p,w^2)$.  
Thus, from now on we assume that  $E = E(\n^2_\n,\n^2_\p,w^2)$.   

With this choice, a variation of $\Lambda$ that leaves the metric 
fixed yields the following:
\beq
    d\Lambda = - (m_\n c^2 + \mun) d\n_\n - (m_\p c^2 + \mup) d\n_\p - 
               \alpha dw^2 \ ,
\eeq
where
\beq
    \mun = {\partial E \over \partial \n_\n} 
           \quad , \quad 
    \mup = {\partial E \over \partial \n_\p} 
           \quad , \quad 
    \alpha = {\partial E \over \partial w^2} \ . 
\eeq
Now the $\K^{\n\p}$, $\K^{\n\n}$, and $\K^{\p\p}$ coefficients of the 
general relativistic formalism are related to the coefficients defined 
above via
\begin{eqnarray}
   \K^{\n\p} &=& {2 c^2 \over \n_\n \n_\p} \alpha \ , \cr
      && \cr
   \K^{\n\n} &=& {m_\n c^2 + \mun \over \n_\n} - {2 c^2 \over \n^2_\n} 
          \left(1 + {w^2 \over 2 c^2}\right) \alpha \ , \cr
      && \cr
   \K^{\p\p} &=& {m_\p c^2 + \mup \over \n_\p} - {2 c^2 \over \n^2_\p} 
          \left(1 + {w^2 \over 2 c^2}\right) \alpha \ .
\end{eqnarray}
In terms of these variables, the pressure $P$ is seen to be
\beq
    P = - E + \mun \n_\n + \mup \n_\p \ .
\eeq

The Newtonian limit of the general relativistic superfluid field 
equations reduce to the following set of 8 equations:
\beq
   0 = {\partial \n_\n \over \partial t} + \partial_i \left(\n_\n 
       v_{\n}^i\right) \ , \label{fullconsvn} 
\eeq
\beq
   0 = {\partial \n_\p \over \partial t} + \partial_i \left(\n_\p 
       v_{\p}^i\right) \ , \label{fullconsvp}
\eeq 
\begin{eqnarray}
   0 &=& {\partial \over \partial t} \left(v_{\n}^i + {2 \alpha \over 
       m_{\n} \n_\n} \left[v_{\p}^i - v_{\n}^i\right]\right) + \cr
     && \cr
     &&v_{\n}^j \partial_j \left(v_{\n}^i + {2 \alpha \over m_{\n} \n_\n} 
       \left[v_{\p}^i - v_{\n}^i\right]\right) + \d^{ij} \partial_j 
       \left(\Phi + {\mu_\n \over m_{\n}}\right) + \cr
     && \cr
     && {2 \alpha \over m_{\n} \n_\n} \d^{ij} \d_{kl} \left(v_{\p}^l - 
         v_{\n}^l\right) \partial_j v_{\n}^k \ , \label{fulleulern}
\end{eqnarray}
and
\begin{eqnarray}
   0 &=& {\partial \over \partial t} \left(v_{\p}^i + {2 \alpha \over 
         m_{\p} \n_\p} \left[v_{\n}^i - v_{\p}^i\right]\right) + \cr
      && \cr
      &&v_{\p}^j \partial_j \left(v_{\p}^i + {2 \alpha \over m_{\p} \n_\p} 
         \left[v_{\n}^i - v_{\p}^i\right]\right) + \d^{ij} \partial_j 
         \left(\Phi + {\mu_\p \over m_{\p}}\right) + \cr
      && \cr 
      &&{2 \alpha \over m_{\p} \n_\p} \d^{ij} \d_{kl} \left(v_{\n}^l - 
         v_{\p}^l\right) \partial_j v_{\p}^k  \ . \label{fulleulerp}
\end{eqnarray}
Finally, the equation for the gravitational potential $\Phi$ is
\beq
    \partial_i \partial^i \Phi = 4 \pi G \left(m_\n \n_\n + m_\p \n_\p
                                 \right) \ . \label{poisson}
\eeq
These equations have been derived independently by Prix \cite{RP01}, 
using a Newtonian Lagrangian-based 
variational principle.

\subsection[]{Comparison with the Formalism of Lindblom and Mendell}

In their studies of oscillating superfluid stars Lindblom and Mendell 
\cite{ML91,M91,LM94,LM95,LM00} use the mass densities, $\rho_{\n}$ 
and $\rho_{\p}$, and the fluid three-``velocities'' $\vec{V}_{\n}$ 
and $\vec{V}_{\p}$ as their main variables.  The latter are not the 
same as the three-velocities used in the main text of this paper, 
rather these ``velocities'' are the ``macroscopically averaged'' 
gradients of the phases of the mesoscopic wave functions that describe 
the superfluid neutrons and the superconducting protons and they are 
actually proportional to the spatial components of our neutron and 
proton momenta.  By comparing our mass-conservation equations with 
those of Lindblom and Mendell, which are given by
\begin{eqnarray}
    0 &=& {\partial \rho_{\n} \over \partial t} + \vec{\nabla} 
          \cdotp \left(\rho_{\n\n} \vec{V}_{\n} + \rho_{\n\p} 
          \vec{V}_{\p}\right) 
          \ , \cr
      && \cr
    0 &=& {\partial \rho_{\p} \over \partial t} + \vec{\nabla} 
          \cdotp \left(\rho_{\p\n} \vec{V}_{\n} + \rho_{\p\p} 
          \vec{V}_{\p}\right) \ ,      
\end{eqnarray}
where $\rho_{\n\n}$, $\rho_{\p\p}$ and $\rho_{\n\p} = \rho_{\p\n}$ 
were defined earlier in the main text, we find that 
\begin{eqnarray}
    \rho_{\n} \vec{v}_{\n} &=& \rho_{\n\n} \vec{V}_{\n} + \rho_{\n\p} 
                               \vec{V}_{\p} \ , \cr
                           && \cr
    \rho_{\p} \vec{v}_{\p} &=& \rho_{\p\n} \vec{V}_{\n} + \rho_{\p\p} 
                               \vec{V}_{\p} \ .
\end{eqnarray}
A final comparison to our fluid momenta yields
\beq
    \alpha = {1 \over 2} {\rho_{\n} \rho_{\p} \over \rho_{\n\p}^2 - 
             \rho_{\n\n} \rho_{\p\p}} \rho_{\n\p} \ .
\eeq
This is a very useful result since it enables us to make contact
with the model that Lindblom and Mendell use to describe the
entrainment effect. 

In Section~III of the main text, we made reference to equations 
(69)-(72) of Lindblom and Mendell \shortcite{LM94}.  Neglecting terms of 
order $(m_e/m_n)^2$, where $m_e$ is the electron mass, then these 
equations are (recalling that $\tilde{\mu}_n = \mu_n/m_n$ etcetera)
\begin{eqnarray}
    \left({\partial \rho \over \partial P} \right)_\beta &=& {\Delta 
    \over \rho} \left[\left({\partial \tilde{\mu}_\n \over \partial 
    \rho_\n}\right)_{\rho_\p} - 2 \left({\partial \tilde{\mu}_\n
    \over \partial \rho_\p}\right)_{\rho_\n} + \right. \cr
    && \cr
    &&\left.\left({\partial \tilde{\mu}_\p \over \partial \rho_\p}
    \right)_{\rho_\n}\right] \ , \label{LMTD1}
\end{eqnarray}
\begin{eqnarray}
    \left({\partial \rho \over \partial \beta} \right)_P &=& {\Delta 
    \over \rho} \left[ \rho_n \left({ \partial \tilde{\mu}_\n \over 
    \partial \rho_\n}\right)_{\rho_\p} - \rho_\p \left({\partial 
    \tilde{\mu}_\p \over \partial \rho_\p}\right)_{\rho_\n} + \right. 
    \cr
    && \cr
    &&\left.(\rho_\p - \rho_\n) \left({\partial \tilde{\mu}_\n \over 
    \partial \rho_\p}\right)_{\rho_\n}\right] \ , \label{LMTD2}
\end{eqnarray}
\begin{eqnarray}
    {\partial \over \partial \beta} \left({\rho_\p \over \rho_\n} 
    \right)_P &=& {\Delta \over \rho} \left[\left({\partial 
    \tilde{\mu}_\n \over \partial \rho_\n}\right)_{\rho_\p} + 2 
    {\rho_\p \over \rho_\n} \left({\partial \tilde{\mu}_\n \over 
    \partial \rho_\p}\right)_{\rho_\n} + \right. \cr
    && \cr
    &&\left. {\rho_\p^2 \over \rho_\n^2} \left({\partial \tilde{\mu}_\p 
     \over \partial \rho_\p}\right)_{\rho_\n}\right] \ , \label{LMTD3}
\end{eqnarray}
and
\beq
    {1 \over \Delta } = \left({\partial \tilde{\mu}_\n \over \partial 
    \rho_\n}\right)_{\rho_\p} \left({\partial \tilde{\mu}_\p \over 
    \partial \rho_\p} \right)_{\rho_\n} - \left({\partial 
    \tilde{\mu}_\n \over \partial \rho_\p}\right)^2_{\rho_\n} \ . 
    \label{LMTD4}
\eeq
Note that these follow by considering $\mu_\n$ and $\mu_\p$ to be 
functions of $\rho_\n$ and $\rho_\p$, and thus re-writing 
(\ref{pressvar}) and (\ref{dbetadef}) in the forms
\beq
    \delta P = (...) \delta \rho_\n + (...) \delta \rho_\p 
               \quad , \quad
    \delta \beta = (...) \delta \rho_\n + (...) \delta \rho_\p \ ,
\eeq
and then inverting to find
\beq
    \delta \rho_\n = (...) \delta P + (...) \delta \beta 
                     \quad , \quad
    \delta \rho_\p = (...) \delta P + (...) \delta \beta \ . 
                     \label{densityvars}
\eeq
Finally, by introducing the various sounds speeds we arrive at Eqns 
(\ref{eqspeed}) - (\ref{Delta}).  

\end{document}